\documentclass[11pt]{article}
\usepackage{amsmath,amssymb,graphicx,setspace,amsthm,float,url}
\usepackage{subcaption}
\usepackage{mathtools,blindtext, geometry}
\usepackage{tabularx}
\usepackage[hidelinks]{hyperref}
\usepackage{algorithm}
\usepackage{algpseudocode}

\theoremstyle{definition}
\newtheorem{definition}{Definition}[section]
\newtheorem{theorem}[definition]{Theorem}
\newtheorem{proposition}[definition]{Proposition}
\newtheorem{lemma}[definition]{Lemma}

\newtheorem{example}[definition]{Example}

\newtheorem{conjecture}[definition]{Conjecture}

\title{\huge \textbf{A new implementation of Network GARCH Model}}
\author{Peiyi Zhou \footnote{Department of Mathematics, Imperial College London, Huxley Building, 180 Queen's Gate, London, SW7 2AZ, U.K.}}
\date{\today}

\providecommand{\keywords}[1]{\par\smallskip\noindent\textbf{Keywords:} #1}

\usepackage{xcolor}
\usepackage{hyperref}

\usepackage{listings}
\newcommand{\codefunc}[1]{\textcolor{blue}{\texttt{#1}}}

\usepackage{caption}
\usepackage{subcaption}

\hypersetup{
    colorlinks,
    linkcolor={blue!50!black},
    citecolor={blue!50!black},
    urlcolor={blue!50!black}
}

\usepackage[round]{natbib}  

\begin{document}
\maketitle

\begin{abstract}
Volatility clustering and spillovers are key features of real-world financial time series when there are a lot of cross-sectional financial assets. While network analysis helps connect stocks that are `similar' or `correlated', which is effective to link volatility spillovers between stocks, contemporary multivariate ARCH-GARCH formulations struggle to represent structured network dependence and remain parsimonious. We introduce the Generalised Network GARCH (GNGARCH) model as a network volatility model that embeds the GARCH dynamics within the Generalised Network Autoregressive (GNAR) framework, to capture the dynamic volatility of financial asset return by both the asset itself and its `neighbouring' assets from the constructed virtual network. The proposed volatility model GNGARCH also addresses the limitations for current studies of network GARCH by adapting neighbouring volatility persistence, dynamic conditional covariance updates, and allowing higher-order neighbouring effects rather than only immediate neighbours. This paper provides the model derivation, vectorisation and conversion, stationarity conditions, and also an extension by incorporating threshold coefficients to capture leverage effects. We show that the GNGARCH is a valid volatility model satisfying the stylised facts of financial return series through simulation. Parameter estimation is then performed by using squared returns as variance proxy and minimising a loss function that is either mean squared error (MSE) or quasi-likelihood (QLIKE). We apply our model on 75 of the most active US stocks under a virtual network, and highlight the model's ability in volatility estimation and forecast.
\end{abstract}


\newpage
\tableofcontents
\clearpage 
\pagestyle{plain}

\newpage
\section{Introduction}
In finance, the term `volatility' is commonly known as the degree of variation of prices disregarding directions, and we usually see the tendency of high- and low-volatility periods to occur in clusters for financial returns. This fundamental phenomenon is named volatility clustering, and related time series models are classified as volatility models.

Classical time series volatility models, most notably the ARCH model \citep{ARCH_intro} and the GARCH extension \citep{GARCH_intro}, have been widely adapted for capturing dynamic volatility in both finance and other interdisciplinary studies. Later, \cite{vec_multivariate_GARCH} and \cite{DCC_GARCH_intro} discussed the extension of univariate ARCH-GARCH to multivariate analysis, in the form of the Baba, Engle, Kraft and Kroner GARCH (BEKK-GARCH) model and the Dynamic Conditional Correlation GARCH (DCC-GARCH) model respectively. These studies have greatly enhanced the applicability of ARCH-GARCH models in multiple dimensions, in the context of many cross-sectional financial stock return series. However, real‐world applications often involve complex interdependencies among variables. Assets being similar (for example, sharing the same or similar sector affiliation and supply-chain exposure) or correlated (showing high correlation in prices/returns) are more likely to influence each other and exhibit stronger volatility spillovers, where shocks propagate normally through constructed network effects rather than isolated pairwise relationships. These contemporary multivariate ARCH-GARCH volatility models mentioned above may not capture such network interactions well, while also suffering severe complexity because many parameters are needed. 

Network analysis therefore offers a parsimonious framework for modelling these complex interdependencies. In recent years, with the development of big data and deep learning methods, researchers have increasingly focused on the integration of network and time series analysis. \cite{network_time_series_intro} first introduced the idea of network time series with a specified network autoregressive (integrated) moving average (NARIMA) model, \cite{network_vector_autoregression_NAR} then proposed the network vector autoregressive (NAR) model, and \cite{GNAR_and_GNAR_package} popularised the generalised network autoregressive (GNAR) model, which combines the classic autoregressive (AR) processes with network structures to provide a novel approach for high‐dimensional network time series analysis. Subsequently, \cite{GNAR_with_COVID19} applied GNAR to analyse COVID-19 hospitalisations, demonstrating its strength in real-world applications and parsimony in model setting. Nonetheless, due to its linearity, GNAR can hardly be seen as a model to investigate the dynamic volatility. 

It is then very natural to embed GARCH within the GNAR framework to implement a network GARCH volatility model to accurately describe both volatilities of asset returns and network dependence. Some existing work, such as the network GARCH model discussed by \cite{current_network_GARCH}, extends GARCH to a reasonable network context, and \cite{Threshold_network_GARCH} further improved the proposed network GARCH model by incorporating the threshold structure on model parameters to model volatility asymmetry and capture leverage effects, while retaining the network framework. These models, although parsimonious, have important limitations, in that they (i) neglect the neighbouring volatility persistence, which includes information on all neighbouring nodes' past volatilities; (ii) only focus on conditional variances based on past information, but drop the covariance updates, which is unrealistic for multivariate analysis; (iii) restrict interactions to first-order neighbours, which are directly connected neighbouring nodes, and therefore omit any possible higher-order neighbouring effects that can affect volatilities along longer network paths. Therefore, this paper aims to implement a new network GARCH volatility model by adapting GNAR concepts, and address the listed limitations of contemporary network GARCH models.

This paper introduces the Generalised Network GARCH (GNGARCH) model in Section \ref{sec: GNGARCH Model and Methods}, with the explicit derivation, model vectorisation, investigation into conversion, and a sketch of stationarity conditions; in accordance with \cite{Threshold_network_GARCH}, we also briefly introduce how to extend our GNGARCH with threshold coefficients. Then, Section \ref{sec: GNGARCH Model Specification and Simulation Results} tests our model via simulated data to show the proposed volatility model is valid for satisfying the stylised facts; meanwhile we describe the practical parameter fitting schemes by numerical optimisation with squared return as the variance proxy, and a loss function that is either mean squared error (MSE) or quasi-likelihood (QLIKE), where the parameter estimates via QLIKE are known as the quasi-maximum likelihood estimates (QMLE). Finally, Section \ref{sec: modelling most-active US stocks} applies the GNGARCH model to a real-world dataset of the most active US stocks to illustrate its ability in volatility estimation and forecast under a constructed virtual network.

\section{GNGARCH Model and Methods}\label{sec: GNGARCH Model and Methods}

Extended from the univariate time series, a network time series can be written as a tuple of a multivariate time series $\mathbf{X}_t = (X_{1, t}, \cdots, X_{d,t})^T$ on a network $\mathcal{G} = (\mathcal{K, E})$ with $\mathcal{K} = \{1, \cdots, d\}$ as the set of total $d$ nodes/vertices and $\mathcal{E}$ as the set of edges. Each single entry $X_{i,t} \in \mathbb{R}$ is the value of univariate time series for node $i$ in the network $\mathcal{G}$ at time $t$. In other words, a node here corresponds to a single time series variable, and edges encode relationships or connections between those variables.

\subsection{Related Network Notions and Notations}
The formulation of a network time series model is driven by its underlying graph topology. Throughout the rest of this paper, we will only consider the undirected, unweighted networks without self‑loops. This simplification is not only analytically trivial, but also reasonable in real financial applications, where we do not believe that stocks influence each other asymmetrically and do not have reliable prior information on connection strengths, so every link is simply treated equally. Furthermore, the absence of self-loops is because each stock's own past volatility is already captured by its own GARCH component, allowing such self-edge would then only duplicate that effect and add no new information. Therefore, we can further simplify key network notions: $r$-stage neighbours, adjacency matrices and connection weights introduced by \cite{network_time_series_intro}, \cite{GNAR_and_GNAR_package}, and \cite{GNAR_with_COVID19}.

\begin{definition}[$r$-stage neighbour]
\label{def: r-stage neighbour}
    We say node $j$ is an $r$-stage neighbour of node $i$ if in a network $\mathcal{G} = (\mathcal{K,E})$, the shortest path connecting node $i$ and node $j$ is of length $r$. Followingly, we write $N_r(i)$ as the set of nodes being as the $r$-stage neighbours of node $i$, and $|N_r(i)|$ represents the total number of $r$-stage neighbours for node $i$.
\end{definition}

\begin{definition}[$r$-stage adjacency matrix]
\label{def: r-stage adjacency matrix}
    We define a matrix $\mathbf{S}_r$ that
    \begin{equation}
        (\mathbf{S}_r)_{ij} = \begin{cases}
            1 &\hbox{if node $j$ is the $r$-stage neighbour of node $i$}\\
            0 &\hbox{o.w.}
        \end{cases}
    \end{equation}
    and such matrix $\mathbf{S}_r$ is known as the $r$-stage adjacency matrix.
\end{definition}
For an undirected network, we always have the symmetry of $\mathbf{S}_r$, examining $i$th row of $\mathbf{S}_r$ directly yields the set of $r$‑stage neighbours of node $i$. Obviously, when $r=1$, our $S_1$ is the commonly used adjacency matrix.

Another important notion describing the network structure would be the connection weights between nodes. Each node connection weight $w_{ij}$ with $i,j \in \mathcal{K}$ lie in $[0,1]$ quantifying `the relevance node $j$ has on node $i$ with respect to neighbourhood regression' \citep{GNAR_with_COVID19}. As we consider the unweighted graph, one way to define the node connection weight $w_{ij}$ would be the reciprocal of the number of $r$‑stage neighbours of $i$, hence each $r$-stage neighbour of node $i$ would be treated in equal importance.

\begin{definition}[Connection weight matrix]
\label{def: connection weight matrix}
    For a network $\mathcal{G} = (\mathcal{K, E})$ with total $d$ nodes and undirected edges, we define the connection weight matrix $\mathbf{W} \in \mathbb{R}^{d\times d}$ that
    \begin{equation}
        w_{ij} = \begin{cases}
            1/|N_r(i)| &\hbox{if node $j$ is the $r$-stage neighbour of node $i$}\\
            0 & \hbox{if node $j$ is not a neighbour of $i$ at any stage}
        \end{cases}
    \end{equation}
\end{definition}
Unlike most weight matrices, the connection weight matrices $\mathbf{W}$ are usually asymmetric, even our network is assumed to be undirected, unweighted and no self-loops.

\subsection{GNGARCH Model}
\label{subsec: GNGARCH Model}
Drawing on the GNAR framework of \cite{GNAR_and_GNAR_package} and \cite{GNAR_with_COVID19}, we replace the AR component with a univariate GARCH process to define our global Generalised Network GARCH (GNGARCH($p, q, [s_1, \cdots, s_q], [r_1, \cdots, r_p]$))\footnote{`Global' means we treat with global parameters $\left[\alpha_0, \{\alpha_k\}_{k=1}^q, \{\gamma_\ell\}_{\ell=1}^p, \{\beta_{kr}\}_{(k=1, r=1)}^{(q, s_k)}, \{\delta_{\ell r'}\}_{(\ell=1, r'=1)}^{(q, r_\ell)}\right]$ across the whole network. It is absolutely possible to consider local parameters depending on the specific nodes that capture the node behaviour more precisely. Due to its complexity, we focus here on the global model only and defer the local-parameter analysis to the Appendices.} in terms of:
\begin{subequations}
\begin{align}
\label{eqn: GNGARCH equation 1}
    \mathbf{X}_t &= \Sigma_t^{1/2}\mathbf{Z}_t \iff \mathbf{X}_{t}\mid\mathcal{F}_{t-1} \sim D(0, \Sigma_t)\\
\label{eqn: GNGARCH equation 2}
    \sigma_{i,t}^2 &= \underbrace{\alpha_0 + \sum_{k=1}^{q}\alpha_k X_{i,t-k}^2 + \sum_{\ell=1}^{p}\gamma_\ell \sigma_{i,t-\ell}^2}_{\text{own GARCH($p,q$) component}} 
    \notag\\
    &\quad + \left[\underbrace{\sum_{k=1}^{q}\sum_{r=1}^{s_k}\beta_{kr} \sum_{j \in N_r(i)}w_{ij}X_{j,t-k}^2}_{\text{neighbouring clustering effect}} + \underbrace{\sum_{\ell=1}^{p} \sum_{r'=1}^{r_\ell}\delta_{\ell r'} \sum_{j \in N_{r'}(i)}w_{ij}\sigma_{j,t-\ell}^2}_{\text{neighbouring persistence effect}}\right]\\
\label{eqn: GNGARCH equation 3}
    \sigma_{ij, t} &= \underbrace{\alpha_0 + \sum_{k=1}^{q}\alpha_k X_{i, t-k}X_{j, t-k} + \sum_{\ell=1}^{p}\gamma_\ell \sigma_{ij, t-\ell}}_{\text{GARCH($p,q$) related, node $(i,j)$ covariance}}
    \notag\\
    &\quad + \frac{1}{2} \sum_{k=1}^{q} \sum_{r=1}^{s_k} \beta_{kr} \left[
     \underbrace{\sum_{\substack{u\in N_r(i)\\u\neq j}} w_{iu}X_{u, t-k}X_{j, t-k}}_{\substack{\text{node $i$'s neighbouring effect}\\
     \text{on clustering w.r.t node $j$}}} + \underbrace{\sum_{\substack{v\in N_r(j)\\v\neq i}} w_{jv}X_{i, t-k}X_{v, t-k}}_{\substack{\text{node $j$'s neighbouring effect}\\
     \text{on clustering w.r.t node $i$}}}\right]
     \notag\\
    &\quad + \frac{1}{2} \sum_{\ell=1}^{p}\sum_{r'=1}^{r_\ell}\delta_{\ell r'} \left[\underbrace{\sum_{\substack{u\in N_{r'}(i)\\ u\neq j}} w_{iu}\sigma_{uj, t-\ell}}_{\substack{\text{node $i$'s neighbouring effect}\\
    \text{on persistence w.r.t node $j$}}} + \underbrace{\sum_{\substack{v\in N_{r'}(i)\\v\neq i}} w_{jv}\sigma_{vi, t-\ell}}_{\substack{\text{node $j$'s neighbouring effect}\\
    \text{on persistence w.r.t node $i$}}}\right]
\end{align}
\end{subequations}
Through the above equations, $\mathbf{X}_t = (X_{1,t},\cdots, X_{d,t})^T$ and $X_{i,t}$ denotes the time series variable observed at node $i$ and time $t$, assumed to have zero conditional mean at all time. In finance, we usually treat $X_{i,t}$ as the return of asset $i$ at time $t$ once each node represents a financial asset, either in the form of simple return or log-return. 

\begin{definition}[Financial returns]
\label{def: financial returns}
    If $\{P_t\}_{t=0}^T$ is a sequence of observed prices for a stock, then for $t=1,\cdots,T$, its simple return and log-return are defined respectively as
    \begin{subequations}
    \begin{align}
    \label{eqn: simple return}
    \text{simple return: } &r_t = \frac{P_t - P_{t-1}}{P_{t-1}}\\
    \label{eqn: log-return}
    \text{log-return: } &r_t = \log P_{t} - \log P_{t-1}
    \end{align}
    \end{subequations}
\end{definition}

$\sigma_{i,t}^2$ is conditional variance of $X_{i,t}$ given all past information $\mathcal{F}_{i,t-1}=\sigma(X_{i,t-1},\ldots,X_{i,0})$, $\sigma_{ij,t}$ is the conditional covariance between $X_{i,t}$ and $X_{j,t}$ conditional on $\mathcal{F}_{i,t-1} \cup \mathcal{F}_{j,t-1}$ (and $\mathcal{F}_{t-1} = \bigcup_{i=1}^d \mathcal{F}_{i,t-1}$), and $\Sigma_t$ is the overall conditional covariance matrix with $(\Sigma_t)_{ii} = \sigma_{i,t}^2$ and $(\Sigma_t)_{ij} = \sigma_{ij, t}$. Lastly, $\mathbf{Z}_t = (Z_{1,t}, \cdots, Z_{d,t})^T$ is a vectorised strict white noise (SWN) variable with zero mean $\mathbb{E}[\mathbf{Z}_{t}] = 0$ and unit variance $\mathrm{var}(\mathbf{Z}_{t}) = \mathbf{I}_d$, where every vectorised variable in the SWN process $\{\mathbf{Z}_{t}\}$ is independent and identically distributed (i.i.d.) across both time $t$ and assets $i$.

As asset volatility is generally known as a latent and unobservable variable, we should apply proxies to invigilate volatility. If we already fit a volatility model, then it is common to use the conditional standard deviation $\sigma_{i,t}$ for volatility as a more direct measure. See more about volatility/variance proxy analysis in Section \ref{subsubsec: volatility/variance proxy}. 

We also assume our model is `decoupled', that is, for \eqref{eqn: GNGARCH equation 2}, our variance term $\sigma_{i,t}^2$ is only made of variance terms $\sigma_{i, t-\ell}^2$ (for the same node), $\sigma_{j, t-\ell}^2$ (for different neighbouring nodes) and squared returns $X_{i, t-k}^2$ (for the same node), $X_{j, t-k}^2$ (for different neighbouring nodes). Likewise, \eqref{eqn: GNGARCH equation 3} also assumes that covariance is only dependent on corresponding node covariates. This parallels the diagonal specification of the bivariate GARCH(1,1) \textit{vec} model without exogenous terms \cite{vec_multivariate_GARCH}, where only the diagonal elements of the coefficient matrices are set to be non-zero, enforcing the `decoupling'.

Similar to the univariate GARCH model, in order to ensure the positivity of conditional variance in \eqref{eqn: GNGARCH equation 2}, we need to put constraints on these global parameters that:
\begin{equation}
\label{eqn: constraints on global parameters}
\alpha_0 > 0, \alpha_k \geq 0, \gamma_\ell \geq 0, \beta_{kr} \geq 0, \delta_{\ell r'} \geq 0
\end{equation}
for $k=1,\cdots,q$, $\ell = 1,\cdots,p$, $r = s_1,\cdots, s_q$ and $r'=r_1, \cdots, r_p$.

Our proposed model also ensures symmetry of covariances in \eqref{eqn: GNGARCH equation 3} that $\sigma_{ij,t} = \sigma_{ji,t}$. However, we generally do not have $\sigma_{i,t}^2 = \sigma_{ii,t}$, which means \eqref{eqn: GNGARCH equation 3} only applies for different nodes' covariances ($i\neq j$), and for variances calculation we should only use \eqref{eqn: GNGARCH equation 2}.

Lastly, note that we include coefficient $1/2$ in \eqref{eqn: GNGARCH equation 3} for the neighbouring clustering and persistence terms. This is based on the assumption that we have equal neighbouring clustering and persistence effects on volatility from node $i$'s $r$-stage neighbours interacting on node $j$, and node $j$'s $r$-stage neighbours interacting on node $i$ respectively. The sum of each node's neighbours would result the overall effects. In contrast, without the $1/2$ coefficient, each pairwise interaction would be counted twice: once from the perspective of node $i$ and once from the perspective of node $j$. By including the coefficient $1/2$, we avoid this double-counting and ensure that the overall influence exerted across the network is appropriately normalised.

\subsubsection{Model Vectorisation}
Since the GNAR framework can be actually treated as a special vectorised AR (VAR) process \citep{GNAR_with_COVID19}, it makes sense to vectorise our GNGARCH formulation using the matrix-vector representation. Since \eqref{eqn: GNGARCH equation 2} and \eqref{eqn: GNGARCH equation 3} are distinct, vectorisation will be applied separately on variance and covariance respectively, and together give a full vectorisation of the covariance matrix $\Sigma_t$. An important lemma we used here is
\begin{lemma}
\label{lemma: Hadamard relation}
    The matrix Hadamard product $\odot$ between node connection weight matrix $\mathbf{W}$ and $r$-stage adjacency matrix $\mathbf{S}_r$ has the following property:
    \begin{equation}
        (\mathbf{W\odot S}_r)_{ij} = w_{ij}(\mathbf{S}_r)_{ij} = \begin{cases}
            w_{ij} &\hbox{if node $j$ is an $r$-stage neighbour of node $i$}\\
            0 &\hbox{o.w.}
        \end{cases}
    \end{equation}
\end{lemma}

\paragraph{Variance vectorisation update}
We may first rewrite \eqref{eqn: GNGARCH equation 2} as
\begin{equation}
\begin{split}
     \sigma_{i,t}^2 = \alpha_0 &+ \sum_{k=1}^{q}\left\{\alpha_k X_{i,t-k}^2 + \sum_{r=1}^{s_k}\beta_{kr} \sum_{j \in N_r(i)}w_{ij}X_{j,t-k}^2\right\}\\
     &+ \sum_{\ell=1}^{p} \left\{\gamma_\ell \sigma_{i,t-\ell}^2 + \sum_{r'=1}^{r_\ell}\delta_{\ell r'} \sum_{j \in N_{r'}(i)}w_{ij}\sigma_{j,t-\ell}^2\right\}
\end{split}
\end{equation}

If we further define $\mathbf{h}_t = (\sigma_{1,t}^2, \cdots, \sigma_{d,t}^2)^T$ separately as the diagonal entries of $\Sigma_t$, then applying Lemma \ref{lemma: Hadamard relation} will give us
\begin{subequations}
\begin{align}
\label{eqn: variance vectorisation deduction 1}
\left[(\mathbf{W \odot S}_r)(\mathbf{X}_t \odot \mathbf{X}_t)\right]_{i} &= \sum_{j=1}^d (\mathbf{W\odot S}_r)_{ij} (\mathbf{X}_t \odot \mathbf{X}_t)_{j} = \sum_{j \in N_r(i)} w_{ij} X_{j,t}^2\\
\label{eqn: variance vectorisation deduction 2}
\left[(\mathbf{W \odot S}_r)\mathbf{h}_t\right]_{i} &= \sum_{j=1}^d (\mathbf{W\odot S}_r)_{ij} (\mathbf{h}_t)_{j} = \sum_{j \in N_r(i)} w_{ij} \sigma_{j, t}^2
\end{align}
\end{subequations}
Therefore, by \eqref{eqn: variance vectorisation deduction 1} and \eqref{eqn: variance vectorisation deduction 2}, we can write the closed form of variance vectorisation as
\begin{align}
\label{eqn: variance vectorisation}
    \mathbf{h}_t &= \alpha_0 \mathbf{1}_d + \sum_{k=1}^q \left[\alpha_k \mathbf{I}_d + \sum_{r=1}^{s_k}\beta_{kr}(\mathbf{W\odot S}_r)\right](\mathbf{X}_{t-k} \odot \mathbf{X}_{t-k})
    \notag\\ 
    &\quad + \sum_{\ell=1}^{p} \left[\gamma_{\ell}\mathbf{I}_d + \sum_{r'=1}^{r_\ell}\delta_{\ell r'}(\mathbf{W\odot S}_{r'})\right]\mathbf{h}_{t-\ell}
    \notag\\
    &= \alpha_0 \mathbf{1}_d + \sum_{k=1}^q \Phi_k (\mathbf{X}_{t-k} \odot \mathbf{X}_{t-k})+ \sum_{\ell=1}^{p} \Theta_\ell \mathbf{h}_{t-\ell}
\end{align}
where respectively we let
\begin{subequations}
\begin{align}
\label{eqn: def. of Phi in vetorisation}
\Phi_k &= \alpha_k \mathbf{I}_d + \sum_{r=1}^{s_k}\beta_{kr}(\mathbf{W\odot S}_r)\\
\label{eqn: def. of Theta in vectorisation}
\Theta_\ell &= \gamma_{\ell}\mathbf{I}_d + \sum_{r'=1}^{r_\ell}\delta_{\ell r'}(\mathbf{W\odot S}_{r'})
\end{align}
\end{subequations}

\paragraph{Covariance vectorisation update}
The covariance vectorisation is less trivial than variance, but we can still apply our definition of $\mathbf{W}, \mathbf{S}_r$ and $\mathbf{X}_t$ to define $\mathrm{diag}(\mathbf{X}_t\mathbf{X}_t^T) = \mathrm{diag}(X_{1,t}^2, \cdots, X_{d,t}^2) \in \mathbb{R}^{d\times d}$ and $\mathrm{diag}(\Sigma_t) = \mathrm{diag}(\sigma_{1,t}^2, \cdots, \sigma_{d,t}^2) \in \mathbb{R}^{d \times d}$. If we denote symmetric $\mathbf{B}_t = \mathbf{X}_t\mathbf{X}_t^T - \mathrm{diag}(\mathbf{X}_t\mathbf{X}_t^T)$ and $\mathbf{D}_t = \Sigma_t - \mathrm{diag}(\Sigma_t)$, then:
\begin{subequations}
\begin{align}
\label{eqn: covariance vectorisation deduction 1}
\left[(\mathbf{W \odot S}_r)\mathbf{B}_t\right]_{ij} &= \sum_{\substack{u\in N_r(i)\\u\neq j}} w_{iu} X_{u,t}X_{j,t}, &\left[\mathbf{B}_t(\mathbf{W \odot S}_r)^T\right]_{ij} &= \sum_{\substack{v\in N_r(j)\\ v\neq i}} w_{jv} X_{i,t}X_{v,t}\\
\label{eqn: covariance vectorisation deduction 2}
\left[(\mathbf{W \odot S}_r)\mathbf{D}_t\right]_{ij} &= \sum_{\substack{u\in N_r(i)\\u\neq j}} w_{iu} \sigma_{uj, t}, &\left[\mathbf{D}_t (\mathbf{W \odot S}_r)^T\right]_{ij} &= \sum_{\substack{v\in N_r(j)\\v\neq i}} w_{jv} \sigma_{vi, t}
\end{align}
\end{subequations}
Then by \eqref{eqn: covariance vectorisation deduction 1} and \eqref{eqn: covariance vectorisation deduction 2}, the covariance vectorisation would be shown as
\begin{equation}
\label{eqn: covariance vectorisation}
\begin{split}
\Sigma_t = \alpha_0 \mathbf{1}_{d\times d} &+ \sum_{k=1}^q \left[\alpha_k \mathbf{X}_{t-k}\mathbf{X}_{t-k}^T + \frac{1}{2}\sum_{r=1}^{s_k}\beta_{kr}\left\{(\mathbf{W \odot S}_r)\mathbf{B}_{t-k} + [(\mathbf{W \odot S}_r)\mathbf{B}_{t-k}]^T\right\}\right]\\
&+ \sum_{\ell=1}^{p} \left[\gamma_\ell \Sigma_{t-\ell} + \frac{1}{2}\sum_{r'=1}^{r_\ell}\delta_{\ell r'}\left\{(\mathbf{W \odot S}_{r'})\mathbf{D}_{t-\ell} + [(\mathbf{W \odot S}_{r'})\mathbf{D}_{t-\ell}]^T\right\}\right]
\end{split}
\end{equation}
where $\mathbf{1}_{d\times d}$ means the $d\times d$ matrix with all entries be 1. The final $\Sigma_t$ is constructed with diagonal entries being as $\mathbf{h}_t$ deduced in \eqref{eqn: variance vectorisation} and all off-diagonal entries via \eqref{eqn: covariance vectorisation}, meaning that we discard diagonal entries of $\Sigma_t$ computed by (\ref{eqn: covariance vectorisation}) and replace them with $\mathbf{h}_t$.

\subsubsection{Model Investigation into Conversion}
\label{subsubsec: model investigation into conversion}
For the univariate case, a GARCH process of $\{X_t\}$ can be equivalently written as an ARMA process to the squared series $\{X_t^2\}$ \citep{GARCH_ARMA_conversion}. This well-known conversion between the univariate GARCH and ARMA allows us to investigate whether our GNGARCH model also admits a VARMA representation.

We will investigate such model conversion by using the component form. We shall define $\eta_{i,t} = X_{i,t}^2 - \sigma_{i,t}^2$. Later in the Appendices we show that $\{\eta_{i,t}\}$ is a white noise (WN) process (under the assumption that we have stationarity for the GNGARCH process and the fourth moment of $X_{i,t}$ is finite), and followingly, with $X_{i,t}^2 - \eta_{i,t} = \sigma_{i,t}^2$, we rewrite the variance part \eqref{eqn: GNGARCH equation 2} as
\begin{equation}
\begin{split}
X_{i,t}^2 - \eta_{i,t} &= \alpha_0 + \sum_{k=1}^{q}\alpha_k X_{i,t-k}^2 + \sum_{\ell=1}^{p}\gamma_\ell (X_{i,t-\ell}^2 - \eta_{i, t-\ell})\\
&\quad + \left[\sum_{k=1}^{q}\sum_{r=1}^{s_k}\beta_{kr} \sum_{j \in N_r(i)}w_{ij}X_{j,t-k}^2 + \sum_{\ell=1}^{p} \sum_{r'=1}^{r_\ell}\delta_{\ell r'} \sum_{j \in N_{r'}(i)}w_{ij}(X_{j, t-\ell}^2 - \eta_{j, t-\ell})\right]\\
\end{split}
\end{equation}
Then we rearrange all squared returns to obtain
\begin{equation}
\label{eqn: GNGARCH variance conversion, component form}
\begin{split}
X_{i,t}^2 &= \alpha_0 + \underbrace{\sum_{k=1}^q \left[\alpha_k X_{i,t-k}^2 + \sum_{r=1}^{s_k}\beta_{kr}\sum_{j\in N_r(i)}w_{ij}X_{j,t-k}^2\right]}_{\text{var. term 1}}\\
&\quad + \underbrace{\sum_{\ell=1}^p \left[\gamma_\ell X_{i,t-\ell}^2 + \sum_{r'=1}^{r_\ell}\delta_{\ell r'}\sum_{j\in N_{r'}(i)}w_{ij}X_{j,t-\ell}^2\right]}_{\text{var. term 2}}\\
&\quad + \eta_{i,t} - \underbrace{\sum_{\ell=1}^p \left[\gamma_\ell \eta_{i, t-\ell} + \sum_{r'=1}^{r_\ell}\delta_{\ell r'}\sum_{j\in N_{r'}(i)}w_{ij}\eta_{j,t-\ell}\right]}_{\text{var. term 3}}
\end{split}
\end{equation}
Based on \eqref{eqn: GNGARCH variance conversion, component form}, we then apply the same vectorisation technique as introduced previously, it is easy for us to deduce
\begin{subequations}
\begin{align}
    \text{var. term 1} &\implies \sum_{k=1}^q \left[\alpha_k \mathbf{I}_d + \sum_{r=1}^{s_k}\beta_{kr}(\mathbf{W \odot S}_r)\right](\mathbf{X}_{t-k}\odot \mathbf{X}_{t-k}) \sim \sum_{k=1}^q \Phi_k \mathbf{X}_{t-k}^2\\
    \text{var. term 2} &\implies \sum_{\ell=1}^p \left[\gamma_\ell \mathbf{I}_d + \sum_{r'=1}^{r_\ell}\delta_{\ell r'}(\mathbf{W \odot S}_{r'})\right](\mathbf{X}_{t-\ell}\odot \mathbf{X}_{t-\ell}) \sim \sum_{\ell=1}^p \Theta_\ell \mathbf{X}_{t-\ell}^2\\
    \text{var. term 3} &\implies \sum_{\ell=1}^p \left[\gamma_\ell \mathbf{I}_d + \sum_{r'=1}^{r_\ell}\delta_{\ell r'}(\mathbf{W \odot S}_{r'})\right]\boldsymbol{\eta}_{t-\ell} \sim \sum_{\ell=1}^p \Theta_\ell \boldsymbol{\eta}_{t-\ell}
\end{align}
\end{subequations}
where we set $\mathbf{X}_t^2 = \mathbf{X}_t \odot \mathbf{X}_t = (X_{1,t}^2, \cdots, X_{d,t}^2)^T$ and $\boldsymbol{\eta}_{t} = (\eta_{1,t}, \cdots, \eta_{d,t})^T$, and the notations of $\Phi_k$ and $\Theta_\ell$ are consistent with \eqref{eqn: def. of Phi in vetorisation} and \eqref{eqn: def. of Theta in vectorisation} respectively. Hence vectorisation on variance term will result a classic VARMA on squared returns as
\begin{equation}
\label{eqn: GNGARCH VARMA variance vec form}
    \mathbf{X}_t^2 = \alpha_0 \mathbf{1}_d + \sum_{m=1}^u \Psi_m \mathbf{X}_{t-m}^2 + \boldsymbol{\eta}_{t} - \sum_{\ell=1}^{p}\Theta_\ell \boldsymbol{\eta}_{t-\ell}
\end{equation}
where $u=\max(p,q)$ and
\begin{equation}
    \Psi_m = \begin{cases}
        \Phi_m + \Theta_m &\hbox{if $1 \leq m \leq \min(p,q)$}\\
        \Phi_m &\hbox{if $\min(p,q) = p < m \leq q$}\\
        \Theta_m &\hbox{if $\min(p,q)=q < m \leq p$}
    \end{cases}
\end{equation}
For the covariance part, we now set $\eta_{ij,t} = X_{i,t}X_{j,t} - \sigma_{ij,t}$ (which can also be shown as a WN process, see the proof in the Appendices), insert this into \eqref{eqn: GNGARCH equation 3} and rearrange similarly to get
\begin{equation}
\label{eqn: GNGARCH covariance conversion, component form}
    X_{i,t}X_{j,t} = \alpha_0 + \text{cov. term 1} + \text{cov. term 2} + \eta_{ij, t} - \text{cov. term 3}
\end{equation}
where we specifically have
\begin{subequations}
\begin{align}
\label{eqn: cov. conversion term 1}
\text{cov. term 1} &= \sum_{k=1}^q \Bigl[\alpha_k\,X_{i,t-k}X_{j,t-k} + \frac{1}{2}\sum_{r=1}^{s_k}\beta_{kr}\Bigl\{\notag\\
&\quad\sum_{\substack{u\in N_r(i)\\u\neq j}} w_{iu}\,X_{u,t-k}X_{j,t-k} + \sum_{\substack{v\in N_r(j)\\v\neq i}} w_{jv}\,X_{i,t-k}X_{v,t-k}\Bigr\}\Bigr]\\
\label{eqn: cov. conversion term 2}
\text{cov. term 2} &= \sum_{\ell=1}^p \Bigl[\gamma_\ell\,X_{i,t-\ell}X_{j,t-\ell} + \frac{1}{2}\sum_{r'=1}^{r_\ell}\delta_{\ell r'}\Bigl\{\notag\\
&\quad\sum_{\substack{u\in N_{r'}(i)\\u\neq j}} w_{iu}\,X_{u,t-\ell}X_{j,t-\ell} + \sum_{\substack{v\in N_{r'}(j)\\v\neq i}} w_{jv}\,X_{i,t-\ell}X_{v,t-\ell}\Bigr\}\Bigr]\\
\label{eqn: cov. conversion term 3}
\text{cov. term 3} &= \sum_{\ell=1}^p \Bigl[\gamma_\ell\,\eta_{ij, t-\ell} + \frac{1}{2}\sum_{r'=1}^{r_\ell}\delta_{\ell r'}\Bigl\{\sum_{\substack{u\in N_{r'}(i)\\u\neq j}} w_{iu}\,\eta_{uj, t-\ell} + \sum_{\substack{v\in N_{r'}(j)\\v\neq i}} w_{jv}\,\eta_{vi, t-\ell}\Bigr\}\Bigr]
\end{align}
\end{subequations}
We now introduce the $vechl$ operator, which operates on a square matrix to give a vector with all lower-triangular entries column-wisely, excluding the diagonal elements. It is trivial to show that $vechl(\mathbf{A+B}) = vechl(\mathbf{A}) + vechl(\mathbf{B})$, then following our previous notation of $\mathbf{B}_t$ and $\mathbf{D}_t$, we will denote
\begin{subequations}
\begin{align}
\label{eqn: def of vechl}
    v_{\mathbf{X}, t} &:= vechl(\mathbf{B}_t) = vechl(\mathbf{X}_t\mathbf{X}_t^T) = (X_{2,t}X_{1,t}, X_{3,t}X_{1,t}, \cdots, X_{d,t}X_{d-1, t})^T\\ 
    v_{\Sigma, t} &:= vechl(\mathbf{D}_t) = vechl(\Sigma_t) = (\sigma_{21, t}, \sigma_{31, t}, \cdots, \sigma_{d(d-1), t})^T\\
    v_{\boldsymbol{\eta}, t} &:= v_{\mathbf{X}, t} - v_{\Sigma, t} = (\eta_{21, t}, \eta_{31, t}, \cdots, \eta_{d(d-1), t})^T
\end{align}
\end{subequations}
For convenience, we introduce two ways of indexing for components of $v_{\mathbf{X},t}, v_{\Sigma,t}$ and $v_{\boldsymbol{\eta},t}$: tuple index and index mapping function.
\begin{definition}[Tuple index]
\label{def: tuple index}
    We name tuple index as an indexing method on the component of $v_{\mathbf{X},t}, v_{\Sigma,t}$ and $v_{\boldsymbol{\eta},t}$ by using a numerical tuple $(m,n)$ that satisfies $1\leq n < m \leq d$, and
    \begin{equation}
        [v_{\mathbf{X},t}]_{(m,n)} = X_{m,t}X_{n,t}
    \end{equation}
    The same tuple index convention is used for $v_{\Sigma, t}$ and $v_{\boldsymbol{\eta}, t}$.
\end{definition}

\begin{definition}[Index mapping function]
\label{def: index mapping function}
    We define bijective $\tau: \{(2,1), \cdots, (d, d-1)\} \to \{1, \cdots, d(d-1)/2\}$ as an index mapping function if it satisfies
    \begin{equation}
        \tau(m,n) = (n-1)d + (m-n) - \frac{n(n-1)}{2}, \quad 1 \leq n < m \leq d
    \end{equation}
\end{definition}
Notably, the index mapping function $\tau$ actually bridges two different forms of indexing for $v_{\mathbf{X},t}, v_{\Sigma,t}$ and $v_{\boldsymbol{\eta},t}$, which
\begin{equation}
    [v_{\mathbf{X},t}]_{(m,n)} = [v_{\mathbf{X},t}]_{\tau(m,n)} = X_{m,t}X_{n,t}
\end{equation}
We now propose an important and intuitive theorem for the possibility of model conversion, and the proof can be found in the Appendices.

\begin{theorem}[Existence of a network-dependent linear transformation]
\label{thm: existence of linear transformation}
With our form of $v_{\mathbf{X}, t}$ as in \eqref{eqn: def of vechl}, for any stage $r$, there always exists a linear transformation matrix $\mathbf{T}_r$ depending on the network connection weights $\mathbf{W}$ that
\begin{equation}
    \mathbf{T}_r=\mathbf{T}_r(\mathbf{W})\in \mathbb{R}^{\frac{d(d-1)}{2} \times \frac{d(d-1)}{2}}
\end{equation}
such that, for all $t$, with index mapping function $\tau$ we always have
\begin{equation}
\label{eqn: proposed linear trans}
\left[\mathbf{T}_r v_{\mathbf{X}, t}\right]_{\tau(i,j)} = \sum_{\substack{u\in N_r(i)\\u\neq j}} w_{iu} X_{u,t-k}X_{j,t-k} + \sum_{\substack{v\in N_r(j)\\v\neq i}} w_{jv} X_{i,t-k}X_{v,t-k}
\end{equation}
\end{theorem}

The above theorem is linear algebraic, which we treat the right hand side of \eqref{eqn: proposed linear trans} as a linear combination of entries of $v_{\mathbf{X},t-k}$ related to the corresponding connection weights. It is trivial that we can replace $v_{\mathbf{X},t}$ with $v_{\Sigma, t}$ or $v_{\boldsymbol{\eta}, t}$ in Theorem \ref{thm: existence of linear transformation}, then
\begin{subequations}
\begin{align}
    \text{cov. term 1} &\implies \sum_{k=1}^q \left[\alpha_k \mathbf{I}_{d(d-1)/2} + \frac{1}{2}\sum_{r=1}^{s_k}\beta_{kr}\mathbf{T}_r\right]v_{\mathbf{X}, t-k} \sim \sum_{k=1}^q \Pi_k v_{\mathbf{X}, t-k}\\
    \text{cov. term 2} &\implies \sum_{\ell=1}^p \left[\gamma_\ell \mathbf{I}_{d(d-1)/2} + \frac{1}{2}\sum_{r'=1}^{r_\ell}\delta_{\ell r'}\mathbf{T}_{r'}\right]v_{\mathbf{X}, t-\ell} \sim \sum_{\ell=1}^p \Lambda_\ell v_{\mathbf{X}, t-\ell}\\
    \text{cov. term 3} &\implies \sum_{\ell=1}^p \left[\gamma_\ell \mathbf{I}_{d(d-1)/2} + \frac{1}{2}\sum_{r'=1}^{r_\ell}\delta_{\ell r'}\mathbf{T}_{r'}\right]v_{\boldsymbol{\eta}, t-\ell} \sim \sum_{\ell=1}^p \Lambda_\ell v_{\boldsymbol{\eta}, t-\ell}
\end{align}
\end{subequations}
where we let
\begin{subequations}
\begin{align}
\label{eqn: def. of Pi in vetorisation}
\Pi_k &= \alpha_k \mathbf{I}_{d(d-1)/2} + \frac{1}{2}\sum_{r=1}^{s_k}\beta_{kr}\mathbf{T}_r\\
\label{eqn: def. of Lambda in vectorisation}
\Lambda_\ell &= \gamma_\ell \mathbf{I}_{d(d-1)/2} + \frac{1}{2}\sum_{r'=1}^{r_\ell}\delta_{\ell r'}\mathbf{T}_{r'}
\end{align}
\end{subequations}
Finally, we obtain the VARMA form on covariance that
\begin{equation}
\label{eqn: GNGARCH VARMA covariance vec form}
    v_{\mathbf{X},t} = \alpha_0\mathbf{1}_{d(d-1)/2} + \sum_{m=1}^{u} \Omega_m v_{\mathbf{X}, t-m} + v_{\eta, t} - \sum_{\ell=1}^p \Lambda_\ell v_{\eta, t-\ell}
\end{equation}
where as before we have $u=\max(p,q)$ and
\begin{equation}
    \Omega_m = \begin{cases}
        \Pi_m + \Lambda_m &\hbox{if $1 \leq m \leq \min(p,q)$}\\
        \Pi_m &\hbox{if $\min(p,q) = p < m \leq q$}\\
        \Lambda_m &\hbox{if $\min(p,q)=q < m \leq p$}
    \end{cases}
\end{equation}

\subsubsection{Stationarity: A Brief Sketch}
Throughout this paper, otherwise stated, `stationarity' is short for the weak/covariance stationarity of the time series. Because our GNGARCH is equivalent to a VARMA and the VARMA stationarity condition coincides with that of the associated VAR, we can apply the GNAR–VAR stationarity results \citep{GNAR_and_GNAR_package} to derive parameter constraints that guarantee stationarity.\footnote{Note that this is a conjecture because such discussion offers merely a sketch of sufficient stationarity conditions based on the variance VARMA form \eqref{eqn: GNGARCH VARMA variance vec form}.  Deriving analogous conditions for the covariance VARMA representation \eqref{eqn: GNGARCH VARMA covariance vec form} is more challenging as the required linear mappings depend intricately on the network structure. However, one might expect that the stationarity constraints for the variance VARMA should be broadly in line with those for the covariance VARMA, so we adapt Conjecture \ref{conj: sufficient stationarity} and employ suitable numerical approximations to ensure the positive-definiteness of $\Sigma_t$ at every $t$. Although it is generally a difficult problem in theory, particularly for large networks \citep{multivariate_GARCH_PD_overview}, practically it can be performed by adding a small `jitter' on the diagonal to force all eigenvalues to be positive, or by projecting onto the nearest positive-definite matrix using, for example, Python’s \codefunc{cov\_nearest} function.}

\begin{conjecture}
\label{conj: sufficient stationarity}
    Let $\mathbf{X}_t$ be a global GNGARCH($p, q, [s_1, \cdots, s_q], [r_1, \cdots, r_p]$) process underlying a network $\mathcal{G} = (\mathcal{K, E})$. If global parameters hold
    \begin{equation}
        \sum_{k=1}^q \left(|\alpha_k| + \sum_{r=1}^{s_k}|\beta_{kr}|\right) + \sum_{\ell=1}^p \left(|\gamma_\ell| + \sum_{r'=1}^{r_\ell}|\delta_{\ell r'}|\right) < 1
    \end{equation}
    together with the non-negativity constraints stated in \eqref{eqn: constraints on global parameters}, then $\mathbf{X}_t$ is stationary.
\end{conjecture}

Because we are not able to provide a general proof of covariance stationarity so far, the stationarity claim is therefore stated as Conjecture \ref{conj: sufficient stationarity}. We will give example of parameters that produce divergent simulation results in Figure \ref{fig: convergent-divergent simulation} in the Appendices. The simulation protocol and further details are described in Section \ref{sec: GNGARCH Model Specification and Simulation Results}.

\subsection{Model Extension: GTN-GARCH}
\label{subsec: model extension: GTN-GARCH}
If we observe or believe, as in many real life cases, the leverage effect, which describes how an asset’s volatility tends to move inversely with its returns, that rising prices usually coincide with lower volatility, and vice versa \citep{leverage}, then it is useful to further enhance our GNGARCH model in Section \ref{subsec: GNGARCH Model} by using volatility asymmetry.

Inspired by univariate threshold GARCH (TGARCH) models \citep{TGARCH} and the recent threshold network GARCH formulation of \cite{Threshold_network_GARCH}, we call our threshold‑augmented framework the Generalised Threshold Network GARCH (GTN‑GARCH) model by incorporating threshold on the node's squared return part. Concretely, one obtains GTN‑GARCH($p, q, [s_1, \cdots, s_q], [r_1, \cdots, r_p]$) by replacing each $\alpha_k$ in \eqref{eqn: GNGARCH equation 2} with $\alpha_k'$ and each $\alpha_k$ in \eqref{eqn: GNGARCH equation 3} with $\alpha_k''$, where
\begin{subequations}
\begin{align}
\label{eqn: GTN-GARCH coef 1}
    \alpha_k' &= \alpha_k^{(+)}\mathbf{1}(X_{i,t-k} \geq 0) + \alpha_k^{(-)}\mathbf{1}(X_{i,t-k}<0)\\
\label{eqn: GTN-GARCH coef 2}
    \alpha_k'' &= \alpha_k^{(+)}\mathbf{1}\left(\min(X_{i,t-k},X_{j,t-k}) \geq 0 \right)
    \notag\\
    &\quad + \alpha_k^{(\mathrm{inter})}\mathbf{1}\left(X_{i,t-k}X_{j,t-k}<0\right)
    \notag\\
    &\quad + \alpha_k^{(-)}\mathbf{1}\left(\max(X_{i,t-k},X_{j,t-k})<0\right)
\end{align}
\end{subequations}

The financial intuition behind the setting of $\alpha_k^{(+)}, \alpha_k^{(-)}, \alpha_k^{(\text{inter})}$ is to explain the leverage effect of financial time series, where we would expect to see the sharpest change of volatility/variance if both of $X_{i,t-k}, X_{j, t-k}$ have negative values (showing a joint crash for both assets); the intermediate change would be the case that only one of them shows negativity corresponding to $\alpha_k^{(\text{inter})}$; for a joint rise (usually indicate a Bull Market), the effect would be measured by $\alpha_k^{(+)}$. GTN-GARCH is definitely a more complicated model, but since we do not change the neighouring effect terms, its statistical properties and vectorisation are similar to those of GNGARCH, and we will give its vectorisation in the Appendices.

\section{GNGARCH Model Specification and Simulation}
\label{sec: GNGARCH Model Specification and Simulation Results}
As \cite{GARCH_intro} pointed out, GARCH$(1,1)$ model can already handle most financial return series without introducing too many parameters from higher-order GARCH, see also \cite{GARCH_ARMA_conversion}. We choose to concentrate on our parsimonious network‐augmented GARCH$(1,1)$ specification, i.e. GNGARCH$(1,1,[1],[1])$, in which each node's conditional variance depends on its own one-period lag and the one-period lag of its first-stage neighbours. With the notation in \eqref{eqn: GNGARCH equation 1}, \eqref{eqn: GNGARCH equation 2} and \eqref{eqn: GNGARCH equation 3}, our parameters for this model are $\boldsymbol{\theta} = (\alpha_0, \alpha_1, \gamma_1, \beta_{11}, \delta_{11})$, which contains significantly fewer parameters than a classic VARMA($1,1$) model, see Table \ref{tab: model parameter number comparison} in the Appendices.

\subsection{Model simulation and validity}
\label{subsec: model simulation and validity}
\begin{figure}[ht]
    \centering
    \includegraphics[width=\linewidth]{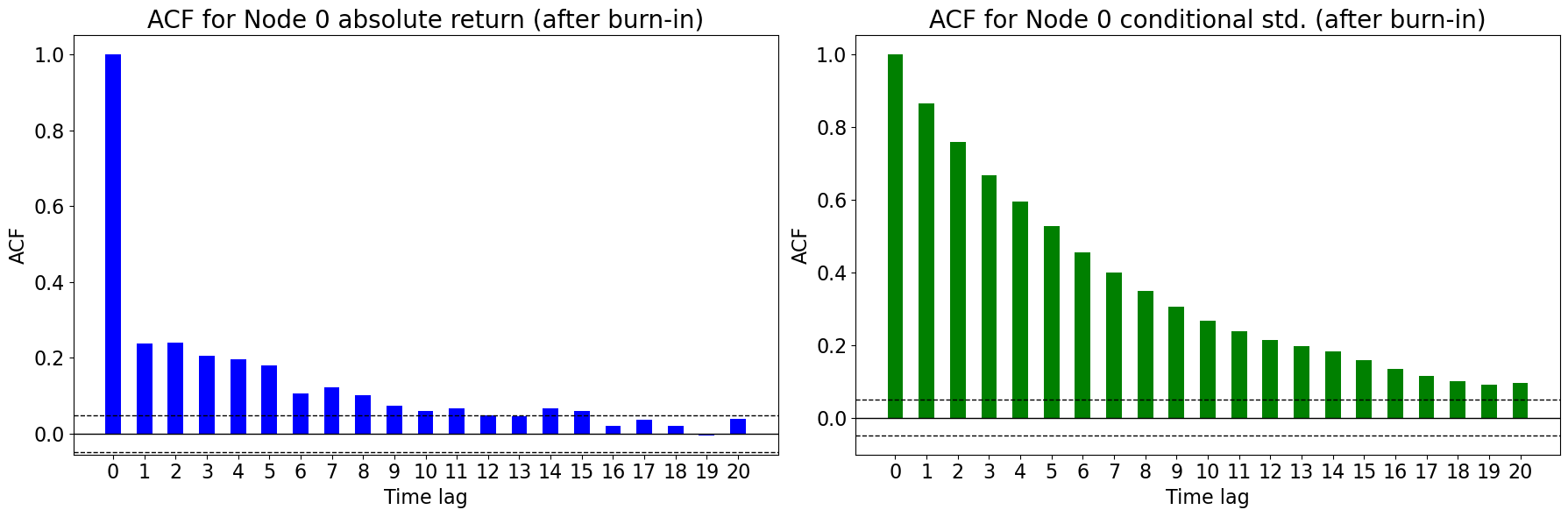}
    \caption{Empirical autocorrelation of the first 20 days of node 0 absolute return (left) and the first 20 lags of simulated conditional standard deviation (right).}
    \label{fig: ACF volatility clustering}
\end{figure}

It is always advisable to analyse the proposed model from the simulated data, where we can then demonstrate the validity of the model by examining the stylised facts of the asset returns, which refers to a range of statistical properties of return series consistently observed across diverse instruments, markets, and time periods \citep{stylised_facts}. In this paper, we will investigate the following stylised facts:
\begin{enumerate}
    \item[(SF1)] The unconditional distribution of returns has a heavy tail and mild asymmetry.
    \item[(SF2)] Autocorrelations of asset returns are often negligible that decay to zero rapidly.
    \item[(SF3)] Volatility clustering and persistence effect: different measures of volatility\footnote{As stated in Section \ref{subsec: GNGARCH Model}, volatility is latent, and we often use a lot of proxies, such as the model-implied conditional standard deviation if we already know the form of the volatility model, or simple realised measures such as the absolute return (see more in later Section \ref{subsubsec: volatility/variance proxy}). The (empirical) autocorrelations for both measures are shown in Figure \ref{fig: ACF volatility clustering}.} shows a positive autocorrelation and slow decay over time lags.
    \item[(SF4)] Aggregational Gaussianity: the distribution of aggregated returns converges toward a normal distribution as the aggregation time scale increases.
    \item[(SF5)] Leverage effect (for GTN-GNGARCH model specifically).
\end{enumerate}

\begin{figure}[ht]
    \centering
    \includegraphics[width=0.6\linewidth]{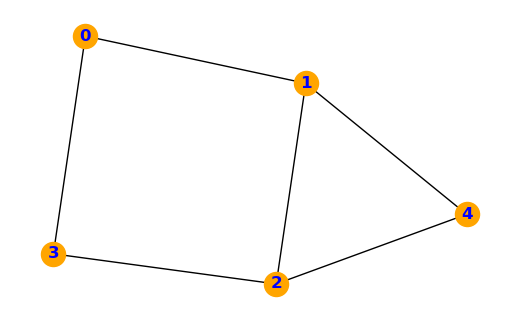}
    \caption{The simple network used for our simulation task and model evaluation. Note that the node index starts from 0 following the Python indexing, hence here we have $\mathbf{X}_t = (X_{0,t}, X_{1,t}, X_{2,t}, X_{3,t}, X_{4,t})^T$.}
    \label{fig: simulation network}
\end{figure}

With the same notation as in Section \ref{subsec: GNGARCH Model}, our simulation of $\{\mathbf{X}_{t}\}$ satisfying GNGARCH model \eqref{eqn: GNGARCH equation 1}, \eqref{eqn: GNGARCH equation 2} and \eqref{eqn: GNGARCH equation 3} will be based on a specific 5 node network shown in Figure \ref{fig: simulation network}, under the assumption that the SWN process $\{\mathbf{Z}_{t}\}$ is normally distributed. In particular, with a preset initial value of $\mathbf{X}_0$ and $\Sigma_0$, the simulation process goes by sequentially update the conditional covariance $\Sigma_t$ via \eqref{eqn: GNGARCH equation 2} and \eqref{eqn: GNGARCH equation 3}, and then simulate $\mathbf{X}_t$ from Gaussian distribution $\mathcal{N}(0, \Sigma_t)$.

We generate a total of 2000 samples of $\mathbf{X}_{t}$ through \eqref{eqn: GNGARCH equation 1}, and discard the first 20\% samples (so 400 burnin samples). The `true' parameters we fit for our GNGARCH$(1,1,[1],[1])$ model for simulation of $\{\mathbf{X}_t\}$ are
\begin{equation}
\label{eqn: model true parameters}
    \alpha_0 = 0.05, \alpha_1 = 0.20, \gamma_1 = 0.60, \beta_{11} = 0.05, \delta_{11} = 0.05
\end{equation}

\begin{figure}[ht]
    \centering
    \includegraphics[width=0.7\linewidth]{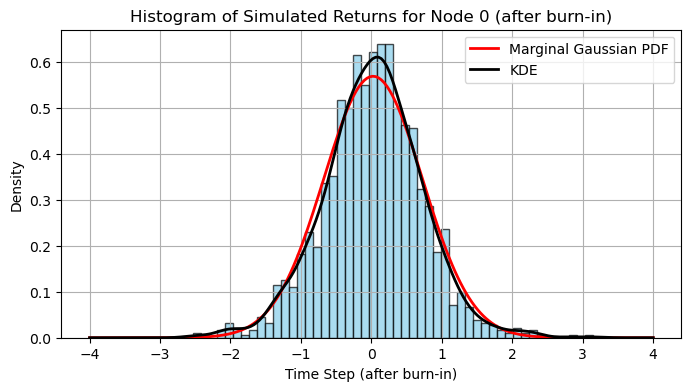}
    \caption{Histogram of node 0 returns with KDE and normal distribution curve.}
    \label{fig: uncon. dist. check}
\end{figure}

For brevity, we present detailed results for node 0, i.e. $\{X_{0,t}\}$, and the remaining nodes exhibit a similar behaviour. We begin with the empirical distribution of the simulated returns. Figure \ref{fig: uncon. dist. check} overlays the return histogram with both a kernel density estimate (KDE) and the Gaussian density fitted using the sample mean and variance. Pearson's kurtosis and simulated series skewness are computed that
\begin{equation}
    \kappa = 3.903 > 3, \quad \beta = 0.008 \neq 0
\end{equation}
A Pearson kurtosis greater than 3 and a non-zero skewness together indicate the heavy-tail and asymmetric nature of simulated return's distribution, aligning with SF1.

\begin{figure}[ht]
    \centering
    \includegraphics[width=.7\linewidth]{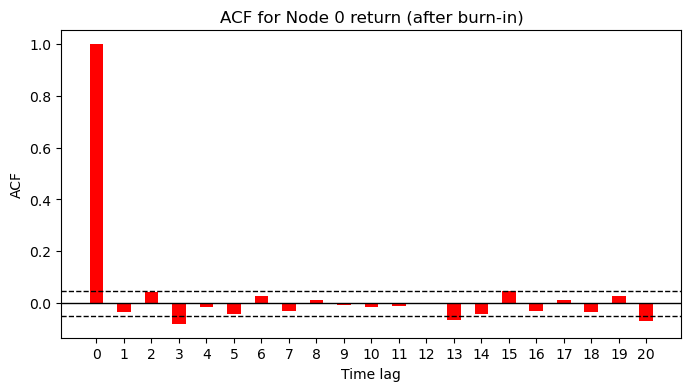}
    \caption{Empirical autocorrelation of the first 20 days (first 20 values) of node 0 return.}
    \label{fig: ACF return uncorrelated} 
\end{figure}

To assess SF2, we compute the sample/empirical autocorrelation function (ACF) of the raw return series
\begin{equation}
    \hat{\rho}_k = \frac{\sum_{t=k+1}^T (X_t - \bar{X})(X_{t-k} - \bar{X})}{\sum_{t=1}^T (X_t - \bar{X})^2}
\end{equation}
for lags $k=1, \cdots, 20$, with the 95\% confidence bound of $\pm 1.96/\sqrt{T}$ (and $T=1600$). We expect to see most $\hat\rho_k$ fall within the bound, with a rapid decay of ACF. Figure \ref{fig: ACF return uncorrelated} exactly explains the desired trend for the first 20 lags of simulated node 0 returns.

We can assess volatility clustering (SF3) either by computing the sample autocorrelation of the absolute return series $|X_{i,t}|$ or by using the sequence of model‑implied conditional standard deviations $\{\sigma_{i,t}\}$ directly. Both measures can be seen as our volatility proxies and calculate their ACF, where the positive ACF values and a slow decay with most of them lying outside the 95\% confidence bound in Figure \ref{fig: ACF volatility clustering} demonstrate both volatility clustering and persistence throughout our GNGARCH volatility model.

\begin{figure}[ht]
    \centering
    \includegraphics[width=\linewidth]{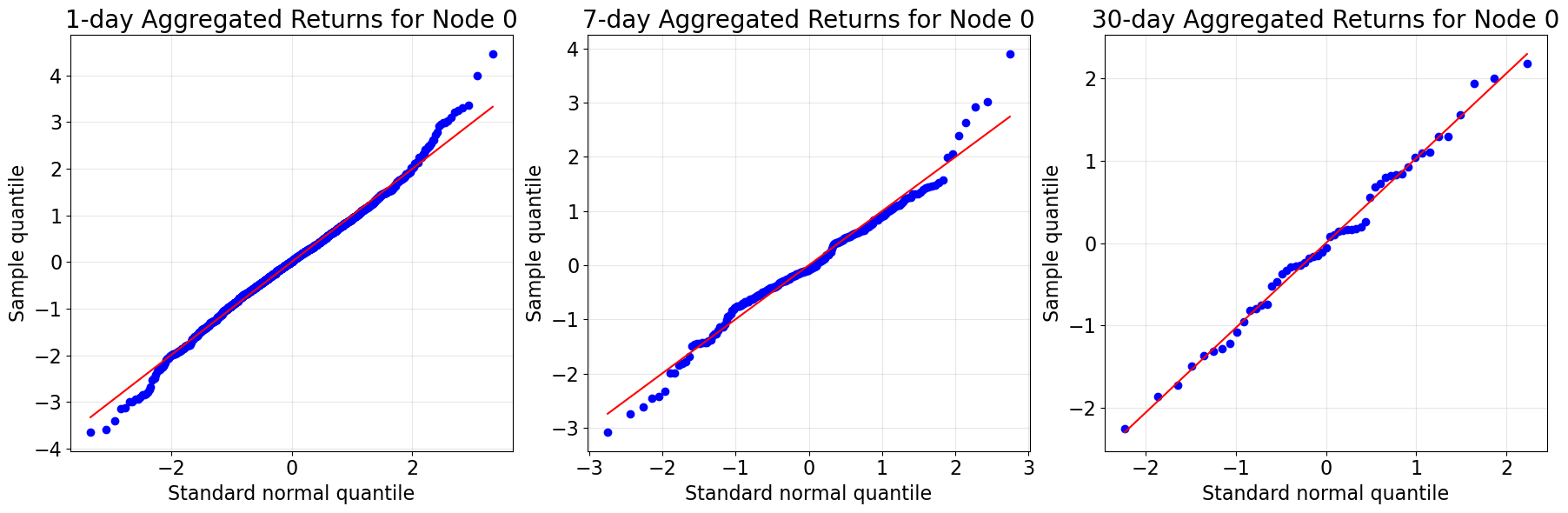}
    \caption{QQ plot for node 0 aggregated returns over days, weeks and months.}
    \label{fig: aggregational Gaussianity SF}
\end{figure}

Examine aggregational Gaussianity (SF4) hinges on having enough observations, which depends on the size of the return series. In our case with a return series of length 1600, it is good to set the aggregation time scale to at most 30 days to balance reasonable sample size and aggregation effects. Followingly we evaluate by considering 1‑day (daily), 7‑day (weekly), and 30‑day (monthly) aggregation windows, and for each we standardise aggregated returns and plot QQ-plots against the standard normal distribution $\mathcal{N}(0,1)$, as shown in Figure \ref{fig: aggregational Gaussianity SF}. As the size of aggregation window increases, fewer points stray from the reference line, showing the desired convergence toward Gaussianity.

\begin{figure}[ht]
\begin{subfigure}{.5\textwidth}
  \centering
  \includegraphics[width=\linewidth]{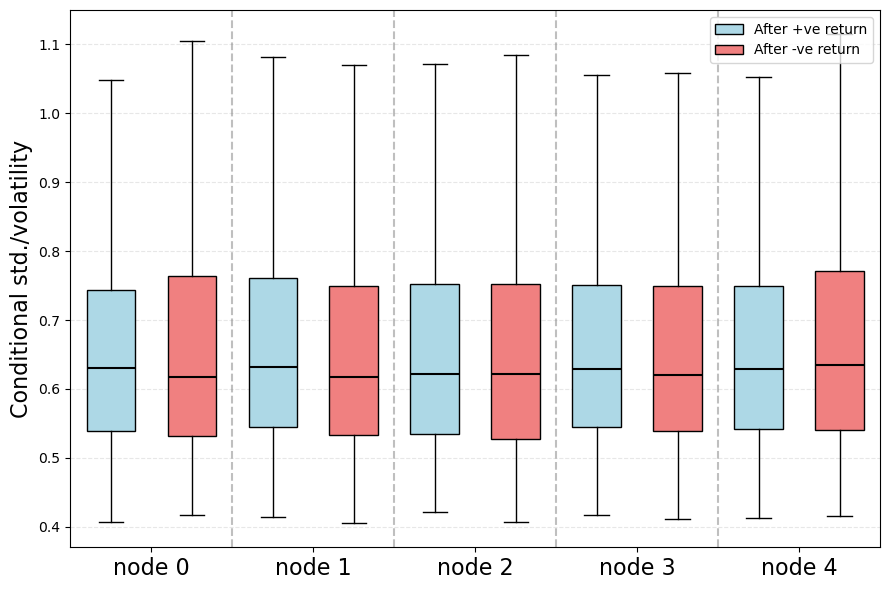}
  \caption{GNGARCH($1,1,[1],[1]$)}
  \label{fig: leverage GNGARCH}
\end{subfigure}%
\begin{subfigure}{.5\textwidth}
  \centering
  \includegraphics[width=\linewidth]{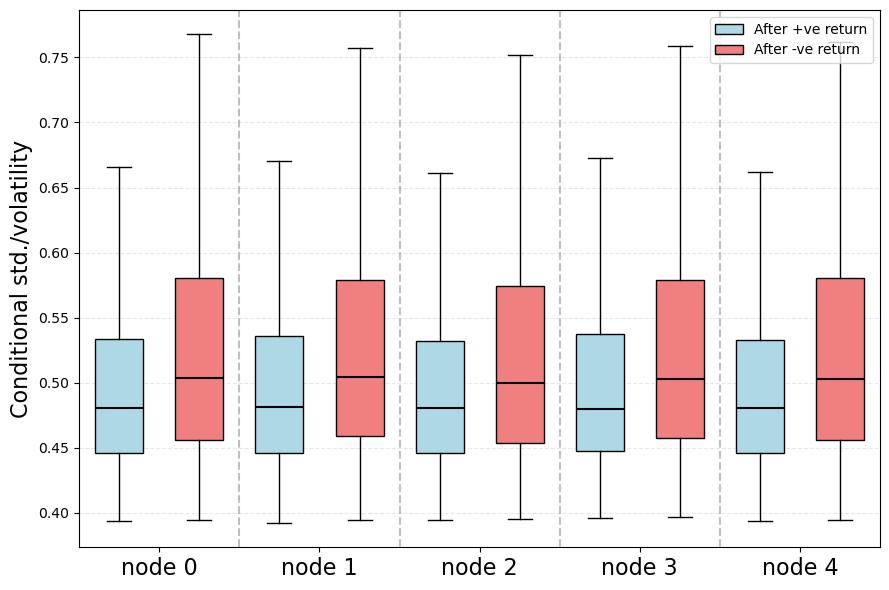}
  \caption{GTN-GNGARCH($1,1,[1],[1]$)}
  \label{fig: leverage GTN-GNGARCH}
\end{subfigure}
\caption{Boxplots of conditional standard deviation across all nodes after a positive/negative shock at most recent time $t-1$. No obvious asymmetry is observed in Figure \ref{fig: leverage GNGARCH} for all nodes in GNGARCH as we do not consider the leverage effect; in contrast, Figure \ref{fig: leverage GTN-GNGARCH} reveals that negative returns result higher subsequent volatility than positive returns (visible for all interquartile range, with a higher value of 25\% quartile, median and 75\% quartile), showing the leverage asymmetry for proposed GTN-GNGARCH model.}
\label{fig: leverage SF}
\end{figure}

We finally illustrate the leverage effect (SF5) by comparison between our GNGARCH and GTN-GNGARCH models. By our representation of models we assume zero conditional mean, so we may simply classify each return as positive or negative to represent price rises and falls, and then examine the subsequent conditional standard deviation as an estimate of volatility. See Figure \ref{fig: leverage SF} for more details.

\subsection{GNGARCH Network Volatility Autocorrelation Function}
Inspired by the Network Autocorrelation Function (NACF) introduced by \cite{GNAR_with_COVID19} which defines in GNAR for graphical aids and quantifies the correlation observed in the network with respect to the $h$-lag, $r$-stage neighbours across all nodes, we can follow this idea to define network volatility autocorrelation function (NVACF) at pair ($h,r$) for our volatility model GNGARCH.

\begin{definition}[Network Volatility Autocorrelation Function (NVACF)]
\label{def: nvacf}
    With connection weight matrix $\mathbf{W}$ and $r$-stage adjacency matrix $\mathbf{S}_r$, for a network with total $d$ nodes, the NVACF of a volatility model GNGARCH is defined in terms of
\begin{equation}
    \mathrm{nvacf}(h,r) = \frac{\sum_{t=1}^{T-h} (\boldsymbol{\sigma}_{t+h}^2 - \overline{\boldsymbol{\sigma}^2})^T (\mathbf{W \odot S}_r + \mathbf{I}_d) (\boldsymbol{\sigma}_{t}^2 - \overline{\boldsymbol{\sigma}^2})}{\sum_{t=1}^{T} (\boldsymbol{\sigma}_{t}^2 - \overline{\boldsymbol{\sigma}^2})^T ((1+\lambda)\mathbf{I}_d) (\boldsymbol{\sigma}_{t}^2 - \overline{\boldsymbol{\sigma}^2})}
\end{equation}
where $\boldsymbol{\sigma}_t^2 = (\sigma_{1,t}^2, \cdots, \sigma_{d,t}^2)^T \in \mathbb{R}^d$ is the vector of conditional variance for all $d$ nodes at time $t$ by \eqref{eqn: GNGARCH equation 2}, and $\overline{\boldsymbol{\sigma}^2} \in \mathbb{R}^d$ is the corresponding vector of sample means (one mean per node) computed across the time dimension, i.e. $(\overline{\boldsymbol{\sigma}^2})_i = \sum_{t=1}^T \sigma_{i,t}^2/T$, with autocovariance bound: $\lambda = \left[\max_{j=1,\cdots, d} \left\{\sum_{i=1}^d [\mathbf{W \odot W}]_{ij}\right\}\right]^{1/2}$.
\end{definition}

\begin{figure}[ht]
    \centering
    \includegraphics[width=.7\linewidth]{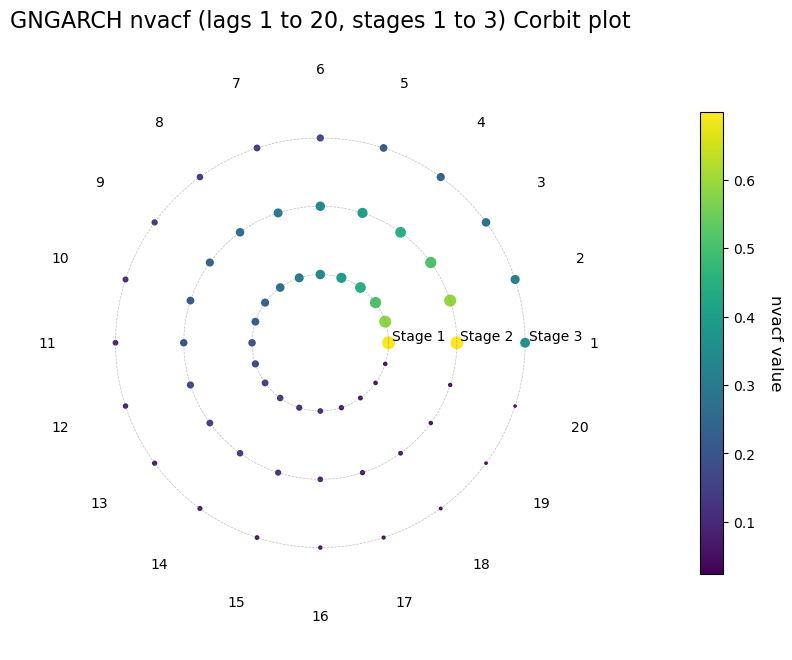}
    \caption{Corbit plot for our simulated data with fitted GNGARCH$(1,1,[1],[1])$ model, up to max lag $h=20$ and max stage of neighbours $r=3$.}
    \label{fig: nvacf corbit plot}
\end{figure}

Analogously to the NACF, our NVACF harnesses the topology of the network, combining the information from each node and its neighbours in stages $r$ to calculate the autocorrelation in lag $h$. On the other hand, NVACF acts as an inferential, model-based plug-in statistic rather than a descriptive one, since volatility is unobservable, and we use the model-implied conditional variance as a proxy to study this latent variable. This makes the NVACF a model-dependent measure that evaluates the network volatility autocorrelation structure implied by the GNGARCH specification. To analyse network autocorrelation decay in a multivariate return series, we first fit the data into the GNGARCH model and then plot the NVACF values via the correlation-orbit (Corbit) plot, shown in Figure \ref{fig: nvacf corbit plot}.

To understand the Corbit plot, numbers around the outermost ring correspond to the time lags $h$, and concentric rings denote the stage of neighbours $r$, where the innermost ring is for the first-stage neighbours, the next ring for the second-stage neighbours, etc. Hence, each point's position encodes both its stage of neighbours $r$ (by which ring it lies on) and its lag $h$ (by its angular placement on that ring). For example, the point on the innermost ring with angular placement at 1 shows $\mathrm{nvacf}(1,1)$, the next anticlockwise point on the same ring is $\mathrm{nvacf}(2,1)$, and the translational point (still with angular placement 1) on the second innermost ring is then $\mathrm{nvacf}(1,2)$.

NVACF shares a similar pattern as the NACF, that correlations decrease both as the time lag $h$ grows and as one moves to more distant $r$-stage neighbours. This aligns with the intuition that volatility shocks lose their influence over time (increase in $h$) and across the network that recent shocks carry stronger autocorrelation, while remote or older disturbances (distant $r$-stage neighbours) have progressively weaker effects.\footnote{This is a common trend over the Corbit plot shown in Figure \ref{fig: nvacf corbit plot}, while we do have observe some exceptions that $\mathrm{nvacf}(h,1) \leq \mathrm{nvacf}(h,2)$ only for $h=1,2,3$. This is because, in our toy network, the counts of first- and second-stage neighbours are comparable, while for most realistic networks, we expect to see that each node's immediate neighbours significantly outnumber neighrbouring nodes at a greater distance, which enforces a clear decline in NVACF as $r$ increases.} Meanwhile, using the autocovariance bound $\lambda$ defined in a similar form as NACF in \cite{GNAR_with_COVID19} also ensures that NVACF values are bounded between $-1$ and $1$, along with $h$ and $r$.

\subsection{Parameter Estimation}
\label{subsec: parameter estimation}
We have so far investigated the model validity via stylised facts and its spatial autocorrelation illustration, indicating a sensible and promising volatility model. An important task for us now is to `re-fit' the parameters with the known simulated returns, which will allow us to complete the parameter fitting of GNGARCH$(1,1,[1],[1])$. Ideally, the recovered estimates should align closely and robustly with their `true' values, indicating an estimate with minimal bias and variance.

Our parameter estimation scheme is based on a numerical optimisation process with a proxy of latent variable volatility and a certain loss function, then employ a gradient‐based optimiser to minimise it. The fitting routine works if the optimiser can recover the ground-truth coefficients on synthetic data. Once validated, we can then apply the same estimation framework to real market return series to complete the parameter calibration of our proposed network-augmented volatility model.

\subsubsection{Volatility/variance Proxy}
\label{subsubsec: volatility/variance proxy}
While the random variable `volatility' is important for capturing the overall trend and pattern of the financial return series and forecasting, it is inherently latent \citep{consistent_ranking_of_vol_models} and cannot be observed directly \citep{GARCH_ARMA_conversion}. People therefore invent different volatility models or proxies to estimate volatility as a form of conditional standard deviation, or sometimes equivalently the conditional variance as its squared form. 

Inevitably, these volatility or variance proxies are only approximations and therefore `imperfect' in essence. In the previous Section \ref{subsec: GNGARCH Model} and \ref{subsec: model simulation and validity}, we discussed using the absolute return as a proxy of volatility, along with the series of conditional standard deviation from the fitted volatility model. By the same logic, we now adapt squared return as a proxy for the true, unobservable conditional variance. The wide use of squared return as the conditional variance proxy is not only convenient, but also its conditionally unbiased nature to the conditional variance in volatility models like GARCH. Accordingly, under our GNGARCH framework, squared returns at time $t$ (vectorised as $\mathbf{X}_t\mathbf{X}_t^T$) are still conditionally unbiased with respect to the model conditional variance $\widehat{\Sigma}_t$.\footnote{To avoid confusion, from here we will replace the use of $\Sigma_t$ into $\widehat{\Sigma}_t$ to distinguish from the true, observed conditional variance $\Sigma_t$ and the GNGARCH model conditional variance $\widehat{\Sigma}_t$, both in vectorised form.} We will include the proof of its conditional unbiasedness in the Appendices.

While both measures are popular and somewhat equivalent, we will prefer model forecasting via conditional variance (so applying the squared return as the proxy) rather than its square-root standard deviation (with absolute return as the proxy). This is because even a conditionally unbiased estimate of the conditional variance will yield a conditionally biased estimate of the conditional standard deviation once you simply take the square root, which is explained in the Appendices by Jensen's inequality. To avoid introducing this systematic distortion, we therefore carry out all comparisons and diagnostics on the model conditional variance for all our later analysis.

Obviously, the squared return is not the only unbiased conditional variance proxy and, although simple, it also suffers from noisy measurements due to the vectorised strict white noise process $\{\mathbf{Z}_t\}$ (where $\mathbf{Z}_t = (Z_{1,t},\cdots, Z_{d,t})^T$). Particularly, for our GNGARCH model, this noise arises because the observed (return) series variable $\mathbf{X}_t$ is generated sequentially by \eqref{eqn: GNGARCH equation 1} through the vectorised strict white noise shocks $\mathbf{Z}_t$, and the subsequent updates of $\Sigma_t$ via \eqref{eqn: GNGARCH equation 2} and \eqref{eqn: GNGARCH equation 3} depend on the previous observations, which are generated by \eqref{eqn: GNGARCH equation 1} with noisy effects beforehand. In other words, noise accumulates across observations, inducing substantial observation‐to‐observation noise relative to the true conditional variance, and consequently worsens the predictive performance of our volatility model \citep{answering_the_skeptics}.

For our network structure, this noise intervention becomes even more pronounced once our network is complex with many nodes, and the measurement noise of each node accumulates throughout the graph. Practitioners sometimes introduce better proxies, such as the realised variance with high-frequency intra-day data, which is a conditional unbiased estimator of the daily conditional variance under the assumptions of zero mean, no jumps, and constant conditional variance over the same day \citep{Patton_2011}. Alternatively, when squared returns remain the proxy of choice, different loss functions in numerical optimization can be applied to mitigate the impact of noise on forecast errors via variance reduction, helping to achieve better forecast performance.

\subsubsection{Loss Function}
\label{subsubsec: loss function}
Another important part of our numerical optimisation is the choice of loss function $L$, which links the chosen variance proxy and the conditional variance derived from the volatility model. In the GNGARCH framework, once we set the squared return $\mathbf{X}_t\mathbf{X}_t^T$ as the variance proxy and $\widehat{\Sigma}_t$ as the model-implied conditional variance, then the loss takes the general form
\begin{equation}
    L(\mathbf{X}_t\mathbf{X}_t^T, \widehat{\Sigma}_t)
\end{equation}
Although numerous loss functions can be used, it is helpful to find losses with some certain good properties. \cite{Patton_2011} rigorously introduced two important features as a good loss: robustness and homogeneity.

\begin{definition}[Robustness]
\label{def: robust loss definition}
    A loss function $L$ is robust if it preserves the ranking of any two (possibly imperfect) volatility/variance forecasts, say $h_{1t}$ and $h_{2t}$, whether one evaluates expected loss against true conditional variance $\sigma_t^2$ or against any unbiased proxy $\tilde{\sigma}_t^2$. Mathematically, for every variance proxy $\tilde{\sigma}_t^2$ satisfying $\mathbb{E}\left(\tilde{\sigma}_t^2\mid\mathcal{F}_{t-1}\right) = \sigma_t^2$, robust loss $L$ ensures
    \begin{equation}
        \mathbb{E}[L(\sigma_t^2, h_{1t})] > \mathbb{E}[L(\sigma_t^2, h_{2t})] \iff \mathbb{E}[L(\hat{\sigma}_t^2, h_{1t})] > \mathbb{E}[L(\hat{\sigma}_t^2, h_{2t})] 
    \end{equation}
    and the above also holds for `$=$' and `$<$' simultaneously.
\end{definition}

The use of robust loss functions leads to robust volatility forecast rankings, meaning they are resistant to noise in the proxy \citep{Patton_2011}. Some may mistakenly believe this property implies immunity to proxy noise; in fact, robust loss only guarantees that forecast rankings are preserved when using the proxy. In other words, using robust loss functions can help to retain the relative rankings of volatility forecast even when the proxies suffer from noise, but it cannot regulate the noise from the proxy itself.

\begin{definition}[Homogeneity]
    A loss function $L$ is homogeneous of order $k$ if
    \begin{equation}
        L(a\tilde{\sigma}_t^2, ah_t) = a^k L(\tilde{\sigma}_t^2, h_t)
    \end{equation}
    for all $a > 0$ and non-negative integers $k$, $\tilde{\sigma}_t$ for the variance proxy and $h_t$ for some imperfect variance forecast (normally the conditional variance from the volatility model).
\end{definition}

\begin{proposition}[\citep{Patton_2011}]
\label{prop: homogeneous loss and rescaling invariant}
    Once the loss is homogeneous, the ranking of any two (imperfect) volatility forecasts by expected loss is invariant to a rescaling of data, while such consistency in ranking via rescaling may not hold if the loss is robust but not homogeneous.
\end{proposition}

Therefore, with the above motivations, we may like to set loss functions that are both robust and homogeneous. \cite{Patton_2011} summarised the form of robust and homogeneous loss functions in detail, and we will focus on the two that are most commonly used: mean squared error (MSE) and quasi-likelihood (QLIKE) function .

\paragraph{MSE}
MSE is regarded as the only loss function that is both robust on the forecast error, set as the difference between variance proxy (using the squared return $\mathbf{X}_t\mathbf{X}_t^T$ here) and the conditional variance of the model $\widehat{\Sigma}_t$, and homogeneous \citep{Patton_2011}. 

Consider we have a sequence of return observations $\{\mathbf{X}_t\}_{t=0}^{T-1}$ (with $\mathbf{X}_0$ as the initial return, and overall we have $T$ data points), and let $\{\widehat{\Sigma}_t\}_{t=1}^{T-1}$ be the corresponding conditional variance sequence from a certain volatility model, that each $\widehat{\Sigma}_t$ is built off information up to $\mathcal{F}_{t-1} = \sigma(\mathbf{X}_0, \cdots, \mathbf{X}_{t-1})$, then for a network with total $d$ nodes, the MSE loss is defined as
\begin{equation}
\label{eqn: MSE}
    L_{\text{MSE}}(\mathbf{X}_t\mathbf{X}_t^T, \widehat{\Sigma}_t) = \frac{1}{T-1}\sum_{t=1}^{T-1} \left[d^{-2}\sum_{i=1}^{d} \sum_{j=1}^{d}(\mathbf{X}_t \mathbf{X}_t^T - \widehat{\Sigma}_t)_{ij}^2\right]
\end{equation}
where the subscript represents the $ij$-th entry of the interested matrix. This measure quantifies the average discrepancy between the squared return proxy and the predicted variances of the model, and we aim to minimise this loss function to capture the accurate volatility forecast from the model.

Although MSE is recognised as a good loss function, which is easy to understand without any further assumptions, also treated as a standard criterion to assess forecasts, its suitability is less straightforward in a non-linear, heteroskedastic setting \citep{answering_the_skeptics}. Many have argued the limitations, and the main problem is that the least-squares estimates can be easily and strongly affected by the outliers of the variance proxy \citep{consistent_ranking_of_vol_models}, where such extreme values again come from the noise effect, as discussed in the previous section \ref{subsubsec: volatility/variance proxy}. Explicitly, outliers of $\mathbf{X}_t\mathbf{X}_t$ due to accumulated noise can severely influence the loss function, result in highly erratic forecast errors, and hence degrade the forecast accuracy.

Quantitatively, \cite{Patton_2011} pointed out that using MSE loss in tuning the volatility model with the usual forecast error can ensure its zero conditional mean (which is good for showing unbiasedness), but its conditional variance is proportional to the square of the variance of the return series. This again highlights the drawbacks of using the squared return as the conditional variance proxy and MSE as the loss function for model parameter estimation.

\paragraph{QLIKE}
Another commonly used parameter estimation scheme is to deduce the maximum likelihood estimator (MLE) via the likelihood function. However, since normally we do not know the exact distribution of the target random variable (for our volatility model, we have no clue about the type of conditional distribution $\mathbf{X_t}\mid\mathcal{F}_{t-1}$) despite knowing its mean and variance, exact likelihood methods are usually infeasible. Statisticians therefore introduce the notion of quasi-maximum likelihood estimator (QMLE) deduced from the quasi-likelihood (QLIKE) function under the assumption that the target variable is under the Gaussian distribution with its mean and variance.

Compared with MSE, although we have assumed Gaussianity for the target variable $\mathbf{X_t}\mid\mathcal{F}_{t-1}$, \cite{QLIKE_consistency_analysis} discussed that we only need that assumption to construct our QLIKE as the loss function, while in volatility models it is not necessary to assume that $\mathbf{X_t}\mid\mathcal{F}_{t-1}$ follows a conditional normal distribution. The resulting QMLE is generally consistent for the parameters analysed: $\boldsymbol{\theta} = (\alpha_0, \alpha_1, \gamma_1, \beta_{11}, \delta_{11})$, and satisfies an asymptotic normality result once standard regularity requirements are met.

The QLIKE function for our GNGARCH volatility model under the Gaussian assumption that $\mathbf{X}_t\mid\mathcal{F}_{t-1} \sim \mathcal{N}(0, \widehat{\Sigma}_t)$, with return sequence $\{\mathbf{X}_t\}_{t=0}^{T-1}$ (with $\mathbf{X}_0$ as the initial return), and conditional variance sequence $\{\widehat{\Sigma}_t\}_{t=1}^{T-1}$, is written as follows.\footnote{While averaging the QLIKE loss over $T-1$ is not strictly necessary, introducing the $1/(T-1)$ factor does not change the location of the minimiser and allows us to treat QLIKE analogously to the averaged MSE in \eqref{eqn: MSE}.  We adapt the same averaging convention when writing the NLL in \eqref{eqn: NLL}.}
\begin{equation}
\label{eqn: QLIKE}
    L_{\text{QLIKE}}(\mathbf{X}_t\mathbf{X}_t^T, \widehat{\Sigma}_t) = \frac{1}{T-1}\sum_{t=1}^{T-1} \left(\log |\widehat{\Sigma}_t| + \mathbf{X}_t^T \widehat{\Sigma}_t^{-1}\mathbf{X}_t \right)
\end{equation}
It is not surprising to see the close relationship between the QLIKE function \eqref{eqn: QLIKE} and the classic negative log-likelihood (NLL) function. Indeed, we show in the Appendices that if we denote the (average) NLL as 
\begin{equation}
\label{eqn: NLL}
    L_{\text{NLL}}(\mathbf{X}_t\mathbf{X}_t^T, \widehat{\Sigma}_t) = -\frac{1}{T-1}\sum_{t=1}^{T-1} \log \mathcal{N}(\mathbf{X}_t; 0, \widehat{\Sigma}_t)
\end{equation}
then the (averaged) QLIKE \eqref{eqn: QLIKE} and NLL \eqref{eqn: NLL} are invariant up to additive constants and an overall scale factor, hence both loss functions will produce the same optimisation result.\footnote{In our Python implementation, we use the \codefunc{nll\_loss} routine. Consequently, our fitting tables and training curves are noted using NLL as the loss function. As QLIKE and NLL are equivalent up to constant and proportionality, we regard them as interchangeable throughout the rest of this paper.} According to \cite{Patton_2011}, QLIKE is the only robust loss function on the standardised forecast error $\mathbf{X}_t^T \widehat{\Sigma}_t^{-1}\mathbf{X}_t$ and is homogeneous. Meanwhile, it is less sensitive to extreme values of $\mathbf{X}_t\mathbf{X}_t^T$, with the conditional variance of the standardised forecast error being approximately a constant, while still guaranteeing the `unbiasedness' behaviour that the conditional expectation of the standardised forecast error equals 1.

\subsubsection{Fitting Results}
\begin{figure}[ht]
    \centering
    \includegraphics[width=\linewidth]{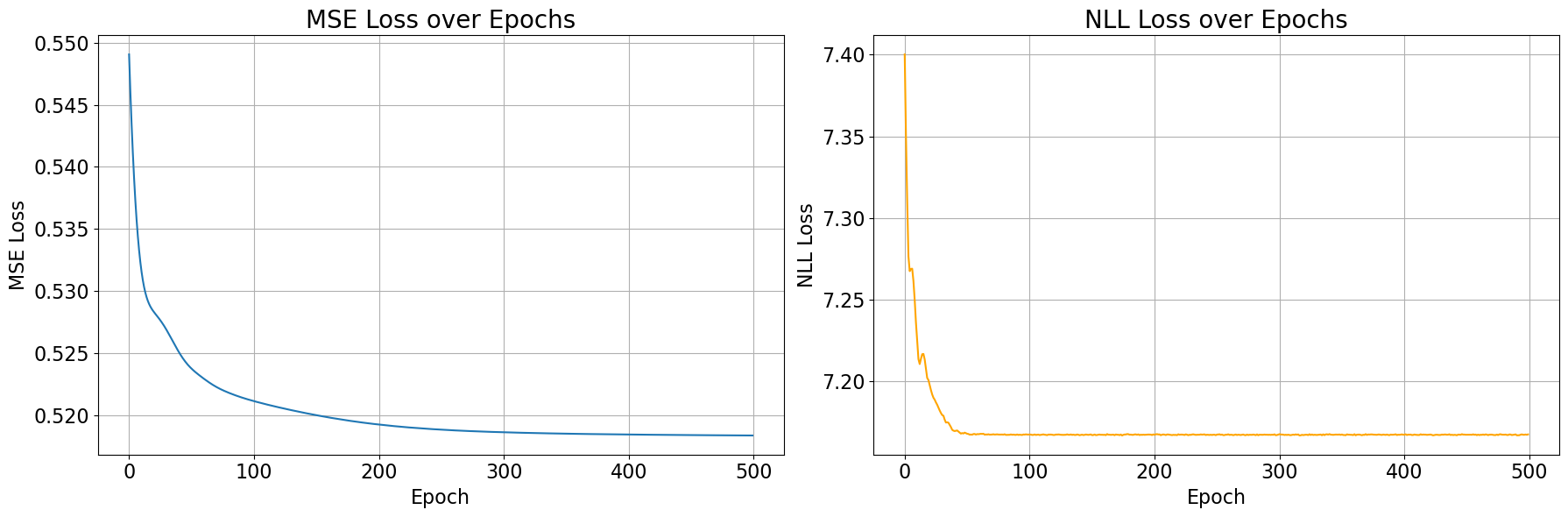}
    \caption{Training curves for MSE loss (left) and NLL loss (right) against epochs.}
    \label{fig: training curve}
\end{figure}

With the above discussion of variance proxy and loss function, we can now carry the optimisation process to fit the specified GNGARCH($1,1,[1],[1]$) model parameters, with the adaptive optimiser \codefunc{Adam} with learning rate 0.01. Table \ref{tab: parameter fitting results} reports both the true parameters (as mentioned in \eqref{eqn: model true parameters}) and the estimated coefficients under both loss functions applied to the simulated return dataset with random seed 0 (which is the same data analysed throughout all previous sections) and 500 epochs, and Figure \ref{fig: training curve} illustrates the convergence of loss over epochs during the training process. Both Table \ref{tab: parameter fitting results} and Figure \ref{fig: training curve} strongly support the validity of our parameter estimation using both loss functions and squared return as the variance proxy, as the estimations are generally close to the true values and the convergence of loss can apparently be observed.

\begin{table}[ht]
\centering
\begin{tabular}{l c r r}
    \hline
    Parameter & True  & Estimate (MSE) & Estimate (NLL) \\
    \hline
    $\alpha_0$ & 0.05 & 0.06 & 0.05\\
    $\alpha_1$ & 0.20 & 0.16 & 0.16\\
    $\gamma_1$ & 0.60 & 0.59 & 0.62\\
    $\beta_{11}$ & 0.05 & 0.03 & 0.05\\
    $\delta_{11}$ & 0.05 & 0.09 & 0.04\\
    \hline
\end{tabular}
\caption{Parameter estimates (2 decimal places) for simulated returns with seed 0 and 500 epochs.}
\label{tab: parameter fitting results}
\end{table}

To further evaluate the robustness of parameter estimates, we re-run the optimisation for small independently simulated datasets (using different random seeds). Here we run 20 replications as they achieve a practical balance between computational cost and reliability of our summary statistics. For each run, we record the estimated coefficients and then summarise, see Table \ref{tab: iterative parameter fitting statistics} and Boxplot (Figure \ref{fig: boxplot parameter fitting}) in the Appendices.

\section{Modelling Most-Active US Stocks}
\label{sec: modelling most-active US stocks}
After analysing the specified GNGARCH($1,1,[1],[1]$) on the simulated data, we are now interested to see whether our model can fit the real return series and make reasonable volatility predictions. The return series dataset used for training the model in this section is taken as log-returns from \eqref{eqn: log-return} of Definition \ref{def: financial returns} derived from the most active 75 US stocks' daily closing price in the market (updated on 19th June, 2025), including well-known stocks like NVIDIA (NVDA), Tesla (TSLA) and Intel (INTC), etc., in a duration from 29th April, 2022 to 31st December, 2024.  

Prior to fitting the GNGARCH$(1,1,[1],[1])$ model to real-world log-returns, we first verify that the log-return series is (first-order) stationary and free from spurious correlations: stationarity ensures we can fit a non-explosive model with appropriate parameters, and no spurious relationship is essential for constructing the virtual network.

\paragraph{Stationarity (first-order)}
Stationarity of the log-returns is crucial, as only then can the proposed GNGARCH($1,1,[1],[1]$) model parameters via the fitting process produce a bounded, well-behaved volatility process in accordance with our theoretical stationarity conditions as in Conjecture \ref{conj: sufficient stationarity}.

Quantitative assessment of stationarity (specifically, first-order stationarity) is often done by the augmented Dickey-Fuller (ADF) test, with the null hypothesis $H_0$ that the process is not stationarity, and the alternative hypothesis $H_1$ that the process is stationary. This test builds upon the original Dickey-Fuller (DF) test, introduced by \cite{DF_test}, which is a unit root test that examines whether an autoregressive (AR) process contains a unit root (a key indicator of non-stationarity). The ADF test extends the basic DF methodology by incorporating more lagged differences of time series into the regression model, see more in \cite{stata_adf}, where such augmentation controls the serial correlation in the data, providing a more powerful test for checking stationarity or trend-stationarity.

The log-return \eqref{eqn: log-return} is in the form of a differencing series; because of this, it is nearly impossible for log-return series to have a trend: a constant increasing trend in log-returns naturally indicates a continuous exponential rise in the price of the stock, and vice versa. For all stocks in our dataset, their log-return series exhibit stationarity. This is confirmed by the ADF tests (without trending), where we reject $H_0$ at the 5\% significance level.

\paragraph{Spurious relationship}
According to \cite{spurious_relationship_dict_def}, a spurious relationship refers to a statistical association between two variables that disappears when controlling confounding variables, which are variables that affect both interested variables (usually known as the exposure and outcome) but do not serve as mediators. The presence of spurious relationships undermines the inferences drawn from correlation-based methods, as significant correlations may emerge solely through confounders, instead of the true relationship between the primary variables.

\cite{Spurious_regresssion} systematically discussed spurious regressions in econometrics and provided a rule for justifying the existence by comparing the $R^2$ and Durbin-Watson statistic ($d$) for any paired variables: when the $R^2 > d$, then it is suspected that there is a spurious relationship between this pair of variables.

With a total of 75 stocks in our dataset, we apply the above diagnostics to check potential spurious relationships across all $\binom{75}{2} = 2775$ stock pairs. For each pair, we first estimate bivariate regressions of log-returns of this stock pair using ordinary least squares (OLS) and then compute $R^2$ and $d$. After verification, there are no pairs satisfying $R^2 > d$, indicating that there is no risk of spurious relationships throughout the dataset.

\subsection{Virtual Network Construction}
We now aim to construct a virtual network for log-returns for all our 75 stocks so that the connection weight matrix $\mathbf{W}$ and $r$-stage adjacency matrix $\mathbf{S}_r$ are deducible. Intuitively, we link two stocks if their (sample) correlation of log-returns is greater than a predetermined threshold. As discussed above, the absence of spurious relationship validates the correlation-based methods for network construction, as the pairwise correlation reflects the true relationship between the log-returns of two stocks.

This correlation-based network construction is also mentioned in \cite{CoC_network_construction}, and the result graph adjacency matrix is named the correlation-of-correlation (CoC) adjacency matrix. Formally, if we denote the empirical sample correlation matrix as $\mathbf{R}$ and the CoC adjacency matrix as $\mathbf{A}$, with threshold $\lambda$, then
\begin{equation}
\label{eqn: CoC adjacency matrix construction}
    \mathbf{A}_{ij} = \mathbf{1}(\mathbf{R}_{ij} > \lambda)
\end{equation}
Since our data are daily log-returns, we may set a monthly window that assumes the log-return for each stock within one month can be seen as i.i.d samples drawn from an identical distribution. Under this assumption, our strategy for constructing the CoC adjacency matrix is then:

\begin{enumerate}
    \item Compute the monthly sample correlation matrix $\mathbf{R}^{(m)}$, that 
    \begin{equation}
        \mathbf{R}^{(m)}_{ij} = \rho(\text{stock}_i^{(m)}, \text{stock}_j^{(m)}) \in \mathbb{R}^{d\times d}
    \end{equation}
    where $\rho$ represents the sample correlation operator, $\text{stock}_i^{(m)}$ is a vector of all daily log-returns for stock $i$ (and $i=1,\cdots, d$ for total $d$ stocks/nodes) within month $m$, for $m=1,\cdots, M$ if there are total $M$ months for our dataset.

    \item Take the absolute value of every element of $\mathbf{R}_{ij}$, as we focus on the strength of the connection rather than the direction, which means that both strong negative and positive correlations indicate strong dependence, vice versa. For convenience, we denote this matrix as $|\mathbf{R}^{(m)}|$ and it is trivially symmetric.

    \item Take the average over months to obtain the integrated correlation matrix $\mathbf{R}$ that
    \begin{equation}
        \mathbf{R} = \frac{1}{M}\sum_{m=1}^M |\mathbf{R}^{(m)}| \in \mathbb{R}^{d\times d}
    \end{equation}

    \item The CoC adjacency matrix $\mathbf{A}$ is then derived by thresholding $\mathbf{R}$ at, for example, the 70\% quantile of all upper off-diagonal values of $\mathbf{R}$\footnote{Only the upper off-diagonal part with total $\binom{d}{2}$ values is needed due to symmetry of $\mathbf{R}$.}, by \eqref{eqn: CoC adjacency matrix construction}.
\end{enumerate}

\begin{figure}[ht]
    \centering
    \includegraphics[width=.9\linewidth]{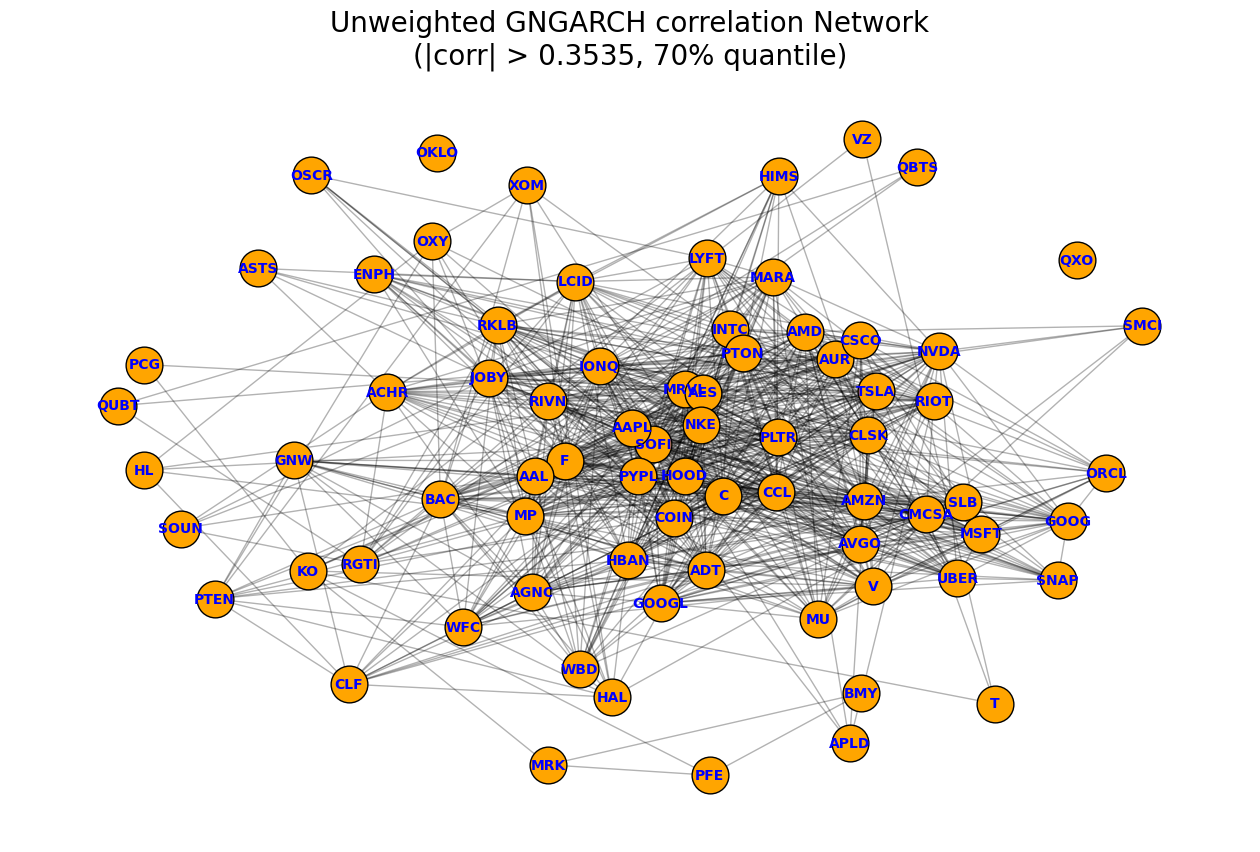}
    \caption{Constructed virtual stock network based on correlation method, with the linking threshold setting as the 70\% quantile of the absolute value of empirical correlation. Node names represent the stock name abbreviations.}
    \label{fig: CoC network}
\end{figure}

The correlation network based on our dataset following the above scheme can be seen in Figure \ref{fig: CoC network}. Although this is definitely not the only sensible way for constructing the virtual network, the correlation network is rather simple and gives us a parsimonious network without requiring any further information.

\subsection{Model Fit, Comparison and Forecast}
With the constructed network, we now apply the fitting scheme in Section \ref{subsec: parameter estimation}, with squared return as the variance proxy and both MSE and QLIKE/NLL losses, to fit a GNGARCH($1,1,[1],[1]$) model on the dataset. Table \ref{tab: stock fitting parameter results} records the parameter estimates.
\begin{table}[ht]
\centering
\begin{tabular}{l r r}
    \hline
    Parameter & Estimate (MSE) & Estimate (NLL)\\
    \hline
    $\alpha_0$ & 0.0232 & 0.0005\\
    $\alpha_1$ & 0.0702 & 0.1648\\
    $\gamma_1$ & 0.1693 & 0.7072\\
    $\beta_{11}$ & 0.0236 & 0.0008\\
    $\delta_{11}$ & 0.0432 & 0.0039\\
    \hline
\end{tabular}
\caption{Parameter estimates on the dataset, results up to 4 decimal places.}
\label{tab: stock fitting parameter results}
\end{table}

In contrast to the close agreement observed in the simulation experiments (Table \ref{tab: parameter fitting results}), the estimates obtained here differ markedly. An interesting point is that, when we change the dataset and resulted network together by excluding stocks whose first recorded price is after 2015 (and there are 47 stocks left), the MSE-based parameter fits show little change, while the QLIKE/NLL-based estimates show a clear mismatch.\footnote{We will include Table \ref{tab: stock fitting parameter results comparison, MSE} and Table \ref{tab: stock fitting parameter results comparison, NLL} in the Appendices for this parameter fitting results and parameter percentage change, as the relative difference (percentage form) between the filtered (47 stock) and full (75 stock) parameter estimates as a degree of mismatch.} This finding is consistent with our earlier discussions in Section \ref{subsec: parameter estimation} that, when the variance proxy is set as the squared return, the MSE-based estimation is heavily affected from the noise effect accumulated in the complex network with many nodes, causing the parameter estimates to be dominated by noise, and thereby shows insensitivity to the sample change.

\begin{figure}[ht]
    \centering
    \includegraphics[width=.9\linewidth]{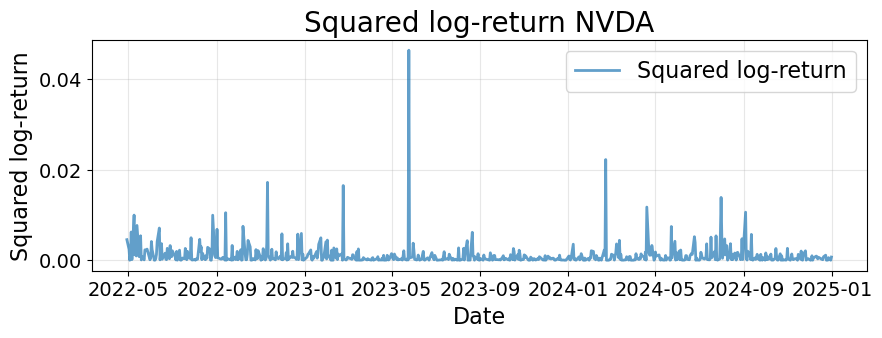}
    \caption{Squared log-returns over time for stock NVDA.}
    \label{fig: squared log returns NVDA}
\end{figure}

We can also show QLIKE/NLL-based parameter fitting performs better by comparing the model conditional variances $\widehat{\Sigma}_t$ with the squared returns. Since squared returns are our variance proxy, we expect they share a similar pattern with model conditional variances. The squared log-return for stock NVDA is shown in Figure \ref{fig: squared log returns NVDA}.

Because our dataset uses daily closing prices, the resulted (log-)returns include overnight jumps and tend to be noisier. To reduce this effect, \cite{rescaling_volatility_forecast} introduced a rescaling technique that matches the level of volatility forecasts to squared returns.\footnote{In \cite{rescaling_volatility_forecast}, the rescaling involves the realised variance, which is defined as the sum of squared intraday returns for high frequency data. However, since our data are daily returns, the realised variance is then approximated as the squared return on this specific day.} We adapt this idea but implement a bit differently for dynamic rescaling, by aligning the level of model conditional variances with observed, centered squared returns, dynamically within a rolling window (with default window size $n_w=252$ for approximate number of trading days in a year). To formulate, for stock $i$, its rescaled conditional variance is
\begin{equation}
    \tilde{\sigma}_{i, t} = c_{i,t}\widehat{\sigma}_{i, t}, \quad     c_{i,t} =  \begin{cases}
    \displaystyle
    \frac{n_w^{-1}\sum_{s=t-n_w+1}^{t} r_{i,s}^2}{n_w^{-1}\sum_{s=t-n_w+1}^t \widehat{\sigma}_{i, s}} &\hbox{when $t\geq n_w$}\\[20pt]
    \displaystyle
    \frac{n_w^{-1}\sum_{s=1}^{n_w} r_{i,s}^2}{n_w^{-1}\sum_{s=1}^{n_w} \widehat{\sigma}_{i, s}} &\hbox{when $t < n_w$}
    \end{cases}
\end{equation}
where $\widehat{\sigma}_{i, t} = ({\widehat{\Sigma}_t})_{ii}$ as the model-implied conditional variance for stock $i$ at time $t$, $r_{i,s}$ is the daily log-return for stock $i$ at time $s$, and $n_w = 252$ for the yearly rolling window. According to Proposition \ref{prop: homogeneous loss and rescaling invariant}, since our MSE and QLIKE loss functions are homogeneous, we still have ranking consistency after rescaling.

After that we can plot and compare the rescaled model conditional variance by using MSE-based parameters and QMLE in Figure \ref{fig: NVDA training volatility} in the Appendices, and QMLE achieves better performance that showing same overall trend as the proxy results in Figure \ref{fig: squared log returns NVDA}.

\begin{figure}[ht]
    \centering
    \includegraphics[width=\linewidth]{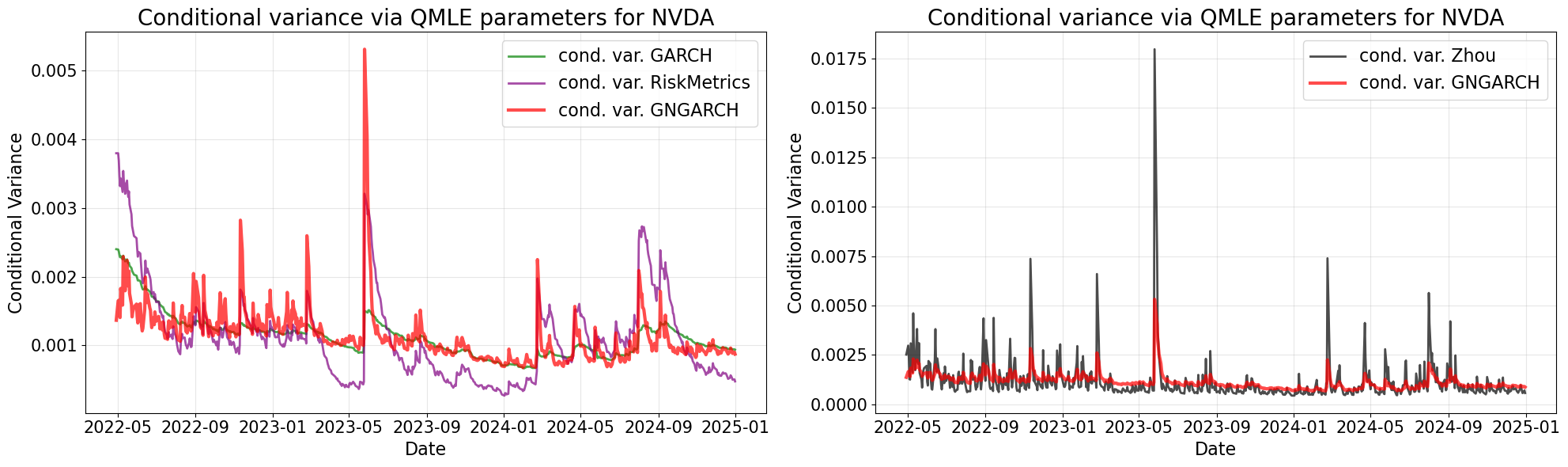}
    \caption{Conditional variance comparison for NVDA across models. Left: comparison between the best-fit univariate GARCH (green), RiskMetrics ($\lambda=0.94$, purple) and the fitted GNGARCH$(1,1,[1],[1])$ (thick red). Right: comparison between Zhou’s network GARCH with QMLE-fitted parameters (black) and the fitted GNGARCH$(1,1,[1],[1])$ (thick red).}
    \label{fig: NVDA volatility comparison}
\end{figure}

We also evaluate the conditional variance for the example stock, NVDA, against four benchmarks: (i) the best-fitting univariate GARCH model; (ii) the RiskMetrics procedure \citep{Patton_2011} defined as $\widehat{\sigma}_{i,t}^2 = \lambda \widehat{\sigma}_{i,t-1} + (1-\lambda)r_{i,t-1}^2$, with $\lambda=0.94$; (iii) our GNGARCH$(1,1,[1],[1])$ estimated by QMLE, and (iv) Zhou’s network GARCH model \citep{current_network_GARCH}, also estimated by QMLE. The results of the comparative conditional variance along with the time for the NVDA stock are shown in Figure \ref{fig: NVDA volatility comparison}.

Figure \ref{fig: squared log returns NVDA} shows intermittent and sharp spikes for NVDA's squared log-returns, particularly around May 2022, September 2022, November 2022, February 2023, May 2023, February 2024, May 2024, August 2024 and September 2024. From the left of Figure \ref{fig: NVDA volatility comparison}, our fitted volatility model GNGARCH($1,1,[1],[1]$) tracks those spikes much more accurately, producing visible and intense local peaks that align with the large, realised spikes; the RiskMetrics provides a much smoother result, and it shows a delay of reaching its peak than the proposed date for the squared log-return, especially after May 2023; the simple, best-fit univariate GARCH leads the smoothest model conditional variance, meaning the model is less reactive and hard to detect the sudden jumps.\footnote{The above discussion is based on NVDA’s log-returns; in the Appendices we present similar plots of model conditional variances and squared log-returns for other stocks: TSLA and AMD, see Figure \ref{fig: squared log-returns TSLA}-\ref{fig: AMD volatility comparison} in the Appendices. Generally, we do not claim that network GARCH models (proposed GNGARCH always outperform the other two methods against the squared (log-)return proxy. However, for a number of stocks, including NVDA, the network-based approaches show closer alignment with the realised squared (log-)returns.}

We then compare our fitted GNGARCH($1,1,[1],[1]$) model against network GARCH model introduced by Zhou \citep{current_network_GARCH}, with fitted parameters $\omega=0.0004, \alpha_0=0.4656, \lambda_0=0.0043, \beta_0=0.4035$ (with the same notation in that paper, up to 4 decimal places). While both network GARCH models capture sharp local peaks coinciding with large realised spikes, Zhou’s model exhibits substantially sharper responses to make our GNGARCH($1,1,[1],[1]$) model appear smoother. Although such discrepancy may be the result of scaling or specification, we are unable to say that GNGARCH$(1,1,[1],[1])$ outperforms Zhou’s model in conditional variance estimation.

\begin{figure}[ht]
    \centering
    \includegraphics[width=\linewidth]{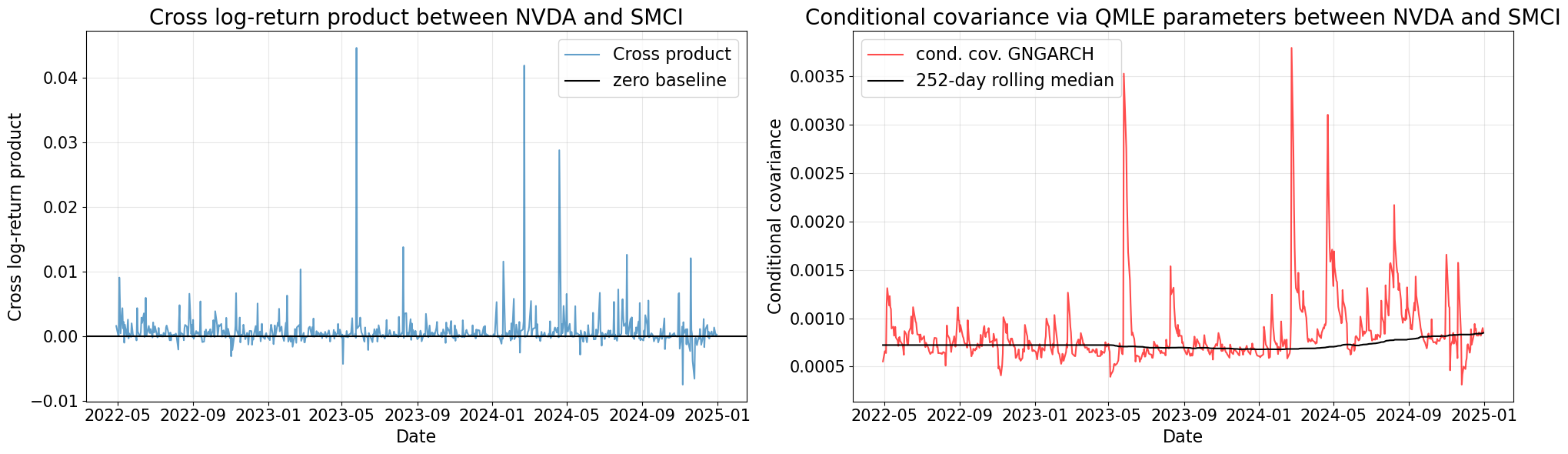}
    \caption{Left: Cross-product of log-returns over time between stock NVDA and SMCI. Right: conditional covariance between NVDA and SMCI, plotted against its (252-day) rolling median. Conditional covariances above the base rolling median indicate the same direction co-movement, vice versa.}
    \label{fig: NVDA-SMCI covolatility}
\end{figure}

On the other hand, the model conditional covariance is often used to estimate the latent (squared) co-volatility \citep{covariance_covolatility}, and the cross-product can be analogously seen as a noisy proxy. Since our GNGARCH includes time-varying covariances, we can investigate pairwise model conditional covariances for two stocks and compare them to the cross-product of corresponding log-returns. We include both in Figure \ref{fig: NVDA-SMCI covolatility}, and comparing the two therefore provides a direct diagnostic of how well the model captures dynamic co-movements. In fact, our fitted GNGARCH($1,1,[1],[1]$) is again able to track the spikes, but also the some troughs that are below the baseline, showing a general consistent trend along with the cross-product of log-returns.

We finally use the constructed network (Figure \ref{fig: CoC network}) and fitted GNGARCH to forecast the stock's volatilities and co-volatilities for the first 6 months of 2025, and we name the corresponding dataset as the validation dataset. Figures \ref{fig: squared log-returns NVDA val}-\ref{fig: NVDA-SMCI covolatility val} in the Appendices display the single stock NVDA's squared log-returns and conditional variance, as well as the NVDA-SMCI log-returns cross-product and conditional covariance. Forecast quality declines somewhat over time, but still roughly shows a similar trend with the squared log-returns and cross-products, confirming GNGARCH's efficacy in joint volatility and co-volatility forecasting in a short period.

\section{Discussion} \label{sec: Conclusion}
This paper develops a novel framework of network GARCH model, the Generalised Network GARCH (GNGARCH), that embeds GARCH dynamics inside the GNAR framework so that an asset's volatility and the pairwise co-volatility with other assets are affected by both its own history and by neighbouring nodes in a constructed network.  Analytical work includes model derivation, vectorisation, conversion with VARMA models and stationarity conditions, and with simulation experiments that on a specified GNGARCH($1,1,[1],[1]$) model and show it is valid that matching the stylised facts of financial return series. Finally, we apply our GNGARCH($1,1,[1],[1]$) on a real-world example on 75 active US stocks, examining and comparing the conditional variance and covariance forecasts for selected stocks with other commonly used volatility models. 

Compared with contemporary studies of network GARCH models, the GNGARCH framework brings two complementary strengths to the literature. It firstly offers a parsimonious way to model a multivariate GARCH process in a network structure, results significantly less model parameters than classic multivariate ARCH-GARCH formulations, such as the BEKK and DCC-GARCH models, or VARMA models. Secondly, it permits higher stage neighbouring effects and dynamic covariance updates, which provides a mechanism for shocks to travel along network paths from different distances, gives us a tool to investigate the time-varying co-volatility under a network structure.

The main limitations of our study of GNGARCH are both methodological and empirical. We do not provide a general proof of stationarity for the covariance part of GNGARCH, but rather stating stationarity as a conjecture. Empirically, we only consider the GNGARCH($1,1,[1],[1]$) specification and do not explore extensions that incorporate higher stage neighbouring nodes. Meanwhile, our model comparisons are qualitative by just plotting conditional variances against squared (log-)returns and covariances against cross-product of corresponding (log-)returns, a quantitative evaluation of predictive accuracy is therefore necessary. Future work will focus on these gaps, along with using the Diebold-Mariano-West tests \citep{DM_test} to quantify predictive accuracy among different volatility models, extend to GTN-GNGARCH, and apply these network GARCH models to asset's value-at-risk (VaR) forecasting.

In conclusion, the GNGARCH offers a parsimonious high-dimensional volatility model with fewer parameters when network relations are meaningful, and gives analysts a tool to trace how volatilities spillovers propagate through the constructed network.

\section{Endmatter}
Codes for this paper are accessible at \url{https://github.com/PZhou114/GNGARCH_coding}, detailed instructions are in the repository's README. For this paper, AI tools like GPT-4o and Grammarly were used to copy-edit the introduction and conclusion text, also improving the quality of written English.

\section*{Acknowledgment}
The author thanks Professor Guy P. Nason for helpful comments on this preprint.

%
%

\clearpage

\bibliographystyle{apalike}
\bibliography{refs}

\clearpage

%
%
%
%

\pagenumbering{arabic}
\renewcommand*{\thepage}{A.\arabic{page}}
\renewcommand{\thesection}{\Alph{section}}
\appendix
\pagebreak

\section{Summary of Notations}
\begin{table}[ht]
\centering
\begin{tabular}{c|c|c}
    \hline
    Symbol & Explanation & Note\\
    \hline
    $\mathcal{G}$ & An (undirected, unweighted) Network & $\mathcal{G} = (\mathcal{K,E})$\\
    $\mathcal{K}$ & Set of nodes/vertices of network $\mathcal{G}$ &\\
    $\mathcal{E}$ & Set of edges of network $\mathcal{G}$ & \\
    $N_r(i)$ & Set of nodes being as the $r$-stage neighbours of node $i$ & see Definition \ref{def: r-stage neighbour}\\
    $\mathbf{S}_r$ & $r$-stage adjacency matrix & see Definition \ref{def: r-stage adjacency matrix}\\
    $\mathbf{W}$ & Connection weight matrix & See Definition \ref{def: connection weight matrix}\\
    $\odot$ & Hadamard product & $(\mathbf{A}\odot\mathbf{B})_{ij} = a_{ij}b_{ij}$\\
    $\sim$ & Behave like, or follow a certain distribution &\\
    $vechl$ & An operator on square matrices to result a vector & see Section \ref{subsubsec: model investigation into conversion}\\
    $[\cdot]_{(m,n)}$ & tuple index & see Definition \ref{def: tuple index}\\
    $\tau(m,n)$ & index mapping function & see Definition \ref{def: index mapping function}\\
    \hline
\end{tabular}
\caption{Summary of commonly-used notations in this paper. Other notations can be located in their corresponding positions with explanations.}
\label{tab: summary of notations}
\end{table}

\newpage
\section{Abbreviations}
\begin{table}[H]
\centering
\begin{tabular}{c|c}
    \hline
    Abbreviation & Explanation\\
    \hline
    ACF & Autocorrelation Function\\
    ADF test & Augmented Dickey-Fuller test\\
    AMD & Advanced Micro Devices, Inc. (AMD), stock name\\
    AR & Autoregressive (process)\\
    ARCH & Autoregressive Conditional Heteroskedasticity (process)\\
    ARMA & Autoregressive Moving Average (process)\\
    BEKK-GARCH & Baba, Engle, Kraft and Kroner GARCH\\
    CoC & Correlation-of-Correlation\\
    DCC-GARCH & Dynamic Conditional Correlation GARCH\\
    GARCH & Generalised ARCH\\
    GNAR & Generalised Network Autoregressive\\
    GNGARCH & Generalised Network GARCH\\
    GTN-GARCH & Generalised Threshold Network GARCH\\
    i.i.d. & independent and identically distributed\\
    KDE & Kernel Density Estimate\\
    MSE & Mean Squared Error\\
    NARIMA & Network Autoregressive Integrated Moving Average\\
    NACF & Network Autocorrelation Function\\
    NLL & Negative log-likelihood\\
    NVACF & Network Volatility Autocorrelation Function\\
    NVDA & NVIDIA Corporation (NVIDIA), stock name\\
    o.w. & otherwise\\
    QLIKE & Quasi-likelihood\\
    QMLE & Quasi Maximum Likelihood Estimator\\
    SF & Stylised Facts (of financial return series)\\
    SMCI & Super Micro Computer, Inc. (Supermicro), stock name\\
    SWN & Strict White Noise\\
    TGARCH & Threshold GARCH\\
    TSLA & Tesla, Inc. (Tesla), stock name\\
    VAR & Vectorised AR (process)\\
    VARMA & Vectorised ARMA (process)\\
    VaR & Value-at-Risk\\
    WN & White Noise\\
    \hline
\end{tabular}
\caption{Abbreviations in this paper, in alphabetical order.}
\label{tab: abbr.}
\end{table}

\newpage
\section{Additional Discussions}
\subsection{Local GNGARCH model representation}
We can definitely change these global parameters to fit each node specifically, which would result the local GNGARCH model that:
\begin{itemize}
    \item Variance term:
    \begin{equation}
    \begin{split}
        \sigma_{i,t}^2 &= \alpha_{i,0} + \sum_{k=1}^{q}\alpha_{i,k} X_{i,t-k}^2+ \sum_{\ell=1}^{p}\gamma_{i,\ell} \sigma_{i,t-\ell}^2\\
        &+ \left[\sum_{k=1}^{q}\sum_{r=1}^{s_k}\beta_{i, kr} \sum_{j \in N_r(i)}w_{ij}X_{j,t-k}^2 +\sum_{\ell=1}^{p} \sum_{r'=1}^{r_\ell}\delta_{i, \ell r'} \sum_{j \in N_{r'}(i)}w_{ij}\sigma_{j,t-\ell}^2\right]
    \end{split}
    \end{equation}
    \item Covariance term:
    \begin{align}
\sigma_{ij, t} &= \alpha_{ij, 0} + \sum_{k=1}^{q}\alpha_{ij, k} X_{i, t-k}X_{j, t-k} + \sum_{\ell=1}^{p}\gamma_{ij, \ell} \sigma_{ij, t-\ell}\notag \\
&+ \Bigl[w^{(ij)} \sum_{k=1}^{q} \sum_{r=1}^{s_k} \beta_{i, kr}
     \sum_{u \in N_r(i), u\neq j} w_{iu}X_{u, t-k}X_{j, t-k}\notag \\
     &\qquad + w^{(ji)}\sum_{k=1}^{q}\sum_{r=1}^{s_k} \beta_{j, kr} \sum_{v \in N_r(j), v\neq i} w_{jv}X_{i, t-k}X_{v, t-k}
\Bigr]\notag \\
&+ \Bigl[w^{(ij)} \sum_{\ell=1}^{p}\sum_{r'=1}^{r_\ell}\delta_{i, \ell r'} \sum_{u \in N_r'(i), u\neq j} w_{iu}\sigma_{uj, t-\ell}\notag \\
    &\qquad + w^{(ji)}\sum_{\ell=1}^{p} \sum_{r'=1}^{r_\ell} \delta_{j, \ell r'}\sum_{v \in N_{r'}(j), v\neq i} w_{jv}\sigma_{iv, t-\ell}\Bigr]
    \end{align}
    where we would have $w^{(ij)} + w^{(ji)} = 1$ for normalisation effect.
\end{itemize}

\subsection{Vectorisation of GTN-GARCH}
With the same notation as covered in \eqref{eqn: variance vectorisation} and \eqref{eqn: covariance vectorisation}, the vectorisation of GTN-GARCH will be
\begin{equation}
\label{eqn: GTN-GARCH variance vectorisation}
\begin{split}
    \mathbf{h}_t = \alpha_0 \mathbf{1}_d &+ \sum_{k=1}^q \left[\mathbf{A}_k + \sum_{r=1}^{s_k}\beta_{kr}(\mathbf{W\odot S}_r)\right](\mathbf{X}_{t-k} \odot \mathbf{X}_{t-k})\\
    &+ \sum_{\ell=1}^{p} \left[\gamma_{\ell}\mathbf{I}_d + \sum_{r'=1}^{r_\ell}\delta_{\ell r'}(\mathbf{W\odot S}_{r'})\right]\mathbf{h}_{t-\ell}
\end{split}
\end{equation}
\begin{equation}
\label{eqn: GTN-GARCH covariance vectorisation}
\begin{split}
\Sigma_t = \alpha_0 \mathbf{1}_{d\times d} &+ \sum_{k=1}^q \left[\mathbf{A}_k' \mathbf{X}_{t-k}\mathbf{X}_{t-k}^T + \frac{1}{2}\sum_{r=1}^{s_k}\beta_{kr}\left\{(\mathbf{W \odot S}_r)\mathbf{B}_{t-k} + [(\mathbf{W \odot S}_r)\mathbf{B}_{t-k}]^T\right\}\right]\\
&+ \sum_{\ell=1}^{p} \left[\gamma_\ell \Sigma_{t-\ell} + \frac{1}{2}\sum_{r'=1}^{r_\ell}\delta_{\ell r'}\left\{(\mathbf{W \odot S}_{r'})\mathbf{D}_{t-\ell} + [(\mathbf{W \odot S}_{r'})\mathbf{D}_{t-\ell}]^T\right\}\right]
\end{split}
\end{equation}
where
\begin{align}
    \mathbf{A}_k &= \alpha_k^{(+)}\mathbf{R}_{t-k} + \alpha_k^{(-)}(\mathbf{I}_d - \mathbf{R}_{t-k})\\
    \mathbf{A}_k' &= \alpha_k^{(+)}\mathbf{P}_{t-k} + \alpha_k^{(\text{inter})}(\mathbf{I}_d - \mathbf{P}_{t-k} - \mathbf{Q}_{t-k}) + \alpha_k^{(-)} \mathbf{Q}_{t-k}\\
    \mathbf{R}_t &= \mathrm{diag}(\mathbf{1}(X_{1, t} \geq 0), \cdots, \mathbf{1}(X_{d, t} \geq 0)) \in \mathbb{R}^{d\times d}\\
    \mathbf{P}_t &\in \mathbb{R}^{d\times d}, \quad (\mathbf{P}_t)_{ij} = \begin{cases}
        \mathbf{1}(\min (X_{i, t}, X_{j,t}) \geq 0) &\hbox{if $i\neq j$}\\
        0 &\hbox{otherwise}
    \end{cases}\\
    \mathbf{Q}_t &\in \mathbb{R}^{d\times d}, \quad (\mathbf{Q}_t)_{ij} = \begin{cases}
        \mathbf{1}(\max (X_{i, t}, X_{j,t}) < 0) &\hbox{if $i\neq j$}\\
        0 &\hbox{otherwise}
    \end{cases}
\end{align}
As usual we only take all off-diagonal entries of \eqref{eqn: GTN-GARCH covariance vectorisation} and replace its diagonal entries with $\mathbf{h}_t$ computed via \eqref{eqn: GTN-GARCH variance vectorisation}.

\section{Proofs and additional Examples}
\begin{proof}
    The following is the proof to show that both $\{\eta_{i,t}\}$ and $\{\eta_{ij,t}\}$ are white noise (WN) processes with $\eta_{i,t} = X_{i,t}^2 - \sigma_{i,t}^2$ and $\eta_{ij,t} = X_{i,t}X_{j,t} - \sigma_{ij,t}$, as described in Section \ref{subsubsec: model investigation into conversion}. Recall that a process $\{X_t\}$ is a WN if it is a covariance stationary process with
    \begin{equation}
        \mathbb{E}(X_t) = 0, \quad \mathrm{var}(X_t)=\sigma^2 < \infty, \quad \mathrm{cov}(X_s, X_t) = 0 \text{ for $s\neq t$}
    \end{equation}
    Now, based on \eqref{eqn: GNGARCH equation 1}-\eqref{eqn: GNGARCH equation 3}, we will have
    \begin{align}
        \mathbb{E}(\mathbf{X}_t\mid\mathcal{F}_{t-1}) = 0 &\implies \mathbb{E}(X_{i,t}\mid\mathcal{F}_{t-1}) = 0\\
        \mathrm{var}(\mathbf{X}_t\mid\mathcal{F}_{t-1}) = \Sigma_t &\implies \begin{cases}
            \mathrm{var}(X_{i,t}\mid\mathcal{F}_{t-1}) = \sigma_{i,t}^2\\
            \mathrm{cov}(X_{i,t}, X_{j,t}\mid\mathcal{F}_{t-1}) = \sigma_{ij,t}
        \end{cases}
    \end{align}
    Also from our expression of \eqref{eqn: GNGARCH equation 2} and \eqref{eqn: GNGARCH equation 3}, we can say that $\sigma_{i,t}^2$ and $\sigma_{ij,t}$ are both $\mathcal{F}_{t-1}$-measurable. This allows us to use Tower property and write
    \begin{align}
        \mathbb{E}(\eta_{i,t}) &= \mathbb{E}(\mathbb{E}(\eta_{i,t}\mid\mathcal{F}_{t-1})) = \mathbb{E}(\mathbb{E}(X_{i,t}^2\mid\mathcal{F}_{t-1}) - \mathbb{E}(\sigma_{i,t}^2\mid\mathcal{F}_{t-1})) = 0\\
        \mathbb{E}(\eta_{ij},t) &= \mathbb{E}(\mathbb{E}(\eta_{ij,t}\mid\mathcal{F}_{t-1})) = \mathbb{E}(\mathbb{E}(X_{i,t}X_{j,t}\mid\mathcal{F}_{t-1}) - \mathbb{E}(\sigma_{ij,t}\mid\mathcal{F}_{t-1})) = 0
    \end{align}
    because of
    \begin{align}
        \mathbb{E}(X_{i,t}^2\mid\mathcal{F}_{t-1}) &= \mathbb{E}(X_{i,t}\mid\mathcal{F}_{t-1})^2 + \mathrm{var}(X_{i,t}\mid\mathcal{F}_{t-1}) = \sigma_{i,t}^2\\
        \mathbb{E}(X_{i,t}X_{j,t}\mid\mathcal{F}_{t-1}) &= \mathbb{E}(X_{i,t}\mid\mathcal{F}_{t-1})\mathbb{E}(X_{j,t}\mid\mathcal{F}_{t-1}) + \mathrm{cov}(X_{i,t}X_{j,t}\mid\mathcal{F}_{t-1}) = \sigma_{ij,t}
    \end{align}
    WLOG we assume $s < t$, and we always have $\eta_{i,s} = X_{i,s}^2 - \sigma_{i,s}^2$ being $\mathcal{F}_{t-1}$-measurable, simply because we trivially have $X_{i,s}$ being $\mathcal{F}_{t-1}$-measurable for $s<t \iff s \leq t-1$, also $\sigma_{i,s}^2$ being $\mathcal{F}_{s-1}$-measurable and $\mathcal{F}_{s-1} \subset \mathcal{F}_{t-1}$, hence $\sigma_{i,s}^2$ is also $\mathcal{F}_{t-1}$-measurable. With the same logic, we can also show that $\eta_{ij,s}$ is $\mathcal{F}_{t-1}$-measurable. Therefore,
    \begin{align}
        \mathrm{cov}(\eta_{i,s}, \eta_{i,t}) &= \mathbb{E}(\eta_{i,s}\eta_{i,t}) = \mathbb{E}[\mathbb{E}(\eta_{i,s}\eta_{i,t}\mid\mathcal{F}_{t-1})] = \mathbb{E}[\eta_{i,s}\mathbb{E}(\eta_{i,t}\mid\mathcal{F}_{t-1})] = 0\\
        \mathrm{cov}(\eta_{ij,s}, \eta_{ij,t}) &= \mathbb{E}(\eta_{ij,s}\eta_{ij,t}) = \mathbb{E}[\mathbb{E}(\eta_{ij,s}\eta_{ij,t}\mid\mathcal{F}_{t-1})] = \mathbb{E}[\eta_{ij,s}\mathbb{E}(\eta_{ij,t}\mid\mathcal{F}_{t-1})] = 0
    \end{align}
    Finally, assume that we have stationarity for the GNGARCH process and the fourth moment of $X_{i,t}$ is finite (and this is usually true in real-world cases), then the variances of $\eta_{i,t}$ and $\eta_{ij,t}$ are also finite.

    When we apply any form of vectorisation later in Section \ref{subsubsec: model investigation into conversion}, the zero mean and zero autocovariance behaviour still holds. For example, in the variance conversion we set $\boldsymbol{\eta}_{t} = (\eta_{1,t}, \cdots, \eta_{d,t})^T$, when $s<t$ we can deduce
    \begin{equation}
        \mathbb{E}(\boldsymbol{\eta}_t) = 0, \quad \mathrm{cov}(\boldsymbol{\eta}_s, \boldsymbol{\eta}_t) = \mathbb{E}(\boldsymbol{\eta}_s\boldsymbol{\eta}_t^T) = \mathbb{E}\begin{pmatrix}
            \eta_{1,s}\eta_{1,t} & \eta_{1,s}\eta_{2,t} & \cdots & \eta_{1,s}\eta_{d,t}\\
            \eta_{2,s}\eta_{1,t} & \eta_{2,s}\eta_{2,t} & \cdots & \eta_{2,s}\eta_{d,t}\\
            \vdots & \vdots & \ddots & \vdots\\
            \eta_{d,s}\eta_{1,t} & \eta_{d,s}\eta_{2,t} & \cdots & \eta_{d,s}\eta_{d,t}
        \end{pmatrix} = \mathbf{0}
    \end{equation}
    This is because $\eta_{i,s}$ is $\mathcal{F}_{t-1}$-measurable, then
    \begin{equation}
        \mathbb{E}(\eta_{i,s}\eta_{j,t}) = \mathbb{E}[\mathbb{E}(\eta_{i,s}\eta_{j,t}\mid\mathcal{F}_{t-1})] = \mathbb{E}[\eta_{i,s}\mathbb{E}(\eta_{j,t}\mid\mathcal{F}_{t-1})] = 0
    \end{equation}
    Under the assumption of stationarity and finiteness of fourth moment of $X_{i,t}$ for all $i=1,\cdots,d$, we also have $\mathrm{var}(\boldsymbol{\eta}_t) = \Omega$ being as a finite and positive definite matrix. Therefore the vectorised form $\boldsymbol{\eta}_t$ is still a vectorised WN process, and we can apply similar discussion to show $v_{\boldsymbol{\eta},t}$ also satisfies WN.
\end{proof}

\begin{proof}
    The following is the constructive proof for Theroem \ref{thm: existence of linear transformation}. Firstly, with $1\leq n < m \leq d$, it is possible for us to use index mapping function (see Definition \ref{def: index mapping function}) and write
    \begin{align}
    \label{proof eqn: thm 2.8 eqn 1}
        \sum_{\substack{u\in N_r(i)\\u\neq j}} w_{iu} X_{u,t}X_{j,t} &= \sum_{m>n} \Bigl\{w_{im}\mathbf{1}(n=j)\mathbf{1}(m\in N_r(i), m\neq j) \notag\\
        &\qquad + w_{in}\mathbf{1}(m=j)\mathbf{1}(n\in N_r(i), n\neq j)\Bigr\}[v_{\mathbf{X},t}]_{\tau(m,n)}\\
        &= \sum_{\tau(m,n)=1}^{\frac{d(d-1)}{2}} \Bigl\{w_{im}\mathbf{1}(n=j)\mathbf{1}(m\in N_r(i), m\neq j) \notag\\
        &\qquad\qquad + w_{in}\mathbf{1}(m=j)\mathbf{1}(n\in N_r(i), n\neq j)\Bigr\}[v_{\mathbf{X},t}]_{\tau(m,n)}
    \end{align}
    and similarly we also have
    \begin{align}
    \label{proof eqn: thm 2.8 eqn 2}
        \sum_{\substack{v\in N_r(j)\\v\neq i}} w_{jv} X_{i,t}X_{v,t} &= \sum_{m>n} \Bigl\{w_{jm}\mathbf{1}(n=i)\mathbf{1}(m\in N_r(j), m\neq i) \notag\\
        &\qquad + w_{jn}\mathbf{1}(m=i)\mathbf{1}(n\in N_r(j), n\neq i)\Bigr\}[v_{\mathbf{X},t}]_{\tau(m,n)}\\
        &= \sum_{\tau(m,n)=1}^{\frac{d(d-1)}{2}} \Bigl\{w_{jm}\mathbf{1}(n=i)\mathbf{1}(m\in N_r(j), m\neq i) \notag\\
        &\qquad\qquad + w_{jn}\mathbf{1}(m=i)\mathbf{1}(n\in N_r(j), n\neq i)\Bigr\}[v_{\mathbf{X},t}]_{\tau(m,n)}
    \end{align}
    Now, we define a non-squared matrix $\widetilde{\mathbf{T}}_r^{(i)} \in \mathbb{R}^{(d-1) \times \frac{d(d-1)}{2}}$, with index mapping function $\tau$ as in Definition \ref{def: index mapping function} that, for the $(j, \tau(m,n))$-th element of $\widetilde{\mathbf{T}}_r^{(i)}$:
    \begin{equation}
        \left[\widetilde{\mathbf{T}}_r^{(i)}\right]_{j, \tau(m,n)} = 
            w_{im}\mathbf{1}(n=j)\mathbf{1}(m\in N_r(i), m\neq j) + w_{in}\mathbf{1}(m=j)\mathbf{1}(n\in N_r(i), n\neq j)
    \end{equation}
    along with $j= \{1,\cdots,d\}\setminus \{i\}$. Then \eqref{proof eqn: thm 2.8 eqn 1} can be vectorised as
    \begin{equation}
    \label{proof eqn: thm 2.8 eqn 3}
        \left[\widetilde{\mathbf{T}}_r^{(i)}v_{\mathbf{X}, t}\right]_j = \sum_{\tau(m,n) = 1}^{\frac{d(d-1)}{2}} \left[\widetilde{\mathbf{T}}_r^{(i)}\right]_{j, \tau(m,n)} [v_{\mathbf{X},t}]_{\tau(m,n)} = \sum_{\substack{u\in N_r(i)\\u\neq j}} w_{iu} X_{u,t}X_{j,t}
    \end{equation}
    and similarly we vectorise \eqref{proof eqn: thm 2.8 eqn 2} as
    \begin{equation}
    \label{proof eqn: thm 2.8 eqn 4}
        \left[\widetilde{\mathbf{T}}_r^{(j)}v_{\mathbf{X}, t}\right]_i = \sum_{\tau(m,n) = 1}^{\frac{d(d-1)}{2}} \left[\widetilde{\mathbf{T}}_r^{(j)}\right]_{i, \tau(m,n)} [v_{\mathbf{X},t}]_{\tau(m,n)} = \sum_{\substack{v\in N_r(j)\\v\neq i}} w_{jv} X_{i,t}X_{v,t}
    \end{equation}
    Unlike traditional indexing, here $j$ is kind of `jump-indexing', which means we will omit the case when $i=j$. In other words, $\widetilde{\mathbf{T}}_r^{(i)}v_{\mathbf{X}, t} \in \mathbb{R}^{(d-1)}$ in jump indexing is 
    \begin{equation}
        \begin{cases}
            \left(\left[\widetilde{\mathbf{T}}_r^{(1)}v_{\mathbf{X}, t}\right]_{2}, \cdots, \left[\widetilde{\mathbf{T}}_r^{(1)}v_{\mathbf{X}, t}\right]_{d}\right)^T &\hbox{$i=1$}\\
            \left(\left[\widetilde{\mathbf{T}}_r^{(d)}v_{\mathbf{X}, t}\right]_{1}, \cdots, \left[\widetilde{\mathbf{T}}_r^{(d)}v_{\mathbf{X}, t}\right]_{d-1}\right)^T &\hbox{$i=d$}\\
            \left(\left[\widetilde{\mathbf{T}}_r^{(i)}v_{\mathbf{X}, t}\right]_{1}, \cdots \left[\widetilde{\mathbf{T}}_r^{(i)}v_{\mathbf{X}, t}\right]_{i-1}, \left[\widetilde{\mathbf{T}}_r^{(i)}v_{\mathbf{X}, t}\right]_{i+1}, \cdots, \left[\widetilde{\mathbf{T}}_r^{(i)}v_{\mathbf{X}, t}\right]_{d}\right)^T &\hbox{o.w.}
        \end{cases}
    \end{equation}
    and we also apply the jump indexing idea on $\widetilde{\mathbf{T}}_r^{(i)}$ as well in terms of
    \begin{align}
        \widetilde{\mathbf{T}}_r^{(1)} &= \begin{pmatrix}
            \left[\widetilde{\mathbf{T}}_r^{(1)}\right]_{2}\\
            \vdots\\
            \left[\widetilde{\mathbf{T}}_r^{(1)}\right]_{d}\\
        \end{pmatrix}, 
        \widetilde{\mathbf{T}}_r^{(d)} = \begin{pmatrix}
            \left[\widetilde{\mathbf{T}}_r^{(d)}\right]_{1}\\
            \vdots\\
            \left[\widetilde{\mathbf{T}}_r^{(d)}\right]_{d-1}\\
        \end{pmatrix},
        \widetilde{\mathbf{T}}_r^{(i)} = \begin{pmatrix}
            \left[\widetilde{\mathbf{T}}_r^{(i)}\right]_{1}\\
            \vdots\\
            \left[\widetilde{\mathbf{T}}_r^{(i)}\right]_{i-1}\\
            \left[\widetilde{\mathbf{T}}_r^{(i)}\right]_{i+1}\\
            \vdots\\
            \left[\widetilde{\mathbf{T}}_r^{(i)}\right]_{d}\\
        \end{pmatrix} \quad \text{for $i\neq 1$}
    \end{align}
    where each $\left[\widetilde{\mathbf{T}}_r^{(i)}\right]_{j} \in \mathbb{R}^{1\times \frac{d(d-1)}{2}}$ represents as the $j$-th row vector of $\widetilde{\mathbf{T}}_r^{(i)}$.
    
    The final step is to use components of non-squared matrices $\widetilde{\mathbf{T}}_r^{(i)}$, and augment into a squared matrix $\mathbf{T}_r \in \mathbb{R}^{\frac{d(d-1)}{2} \times \frac{d(d-1)}{2}}$, which can be done by using jump tuple index and a summation of two parts that
    \begin{equation}
        \mathbf{T}_r = \begin{pmatrix}
            \left[\widetilde{\mathbf{T}}_r^{(2)}\right]_{1}\\[6pt]
            \left[\widetilde{\mathbf{T}}_r^{(3)}\right]_{1}\\[6pt]
            \vdots\\[6pt]
            \left[\widetilde{\mathbf{T}}_r^{(d)}\right]_{1}\\[6pt]
            \left[\widetilde{\mathbf{T}}_r^{(3)}\right]_{2}\\[6pt]
            \vdots\\[6pt]
            \left[\widetilde{\mathbf{T}}_r^{(d)}\right]_{d-1}
        \end{pmatrix} + \begin{pmatrix}
            \left[\widetilde{\mathbf{T}}_r^{(1)}\right]_{2}\\[6pt]
            \left[\widetilde{\mathbf{T}}_r^{(1)}\right]_{3}\\[6pt]
            \vdots\\[6pt]
            \left[\widetilde{\mathbf{T}}_r^{(1)}\right]_{d}\\[6pt]
            \left[\widetilde{\mathbf{T}}_r^{(2)}\right]_{3}\\[6pt]
            \vdots\\[6pt]
            \left[\widetilde{\mathbf{T}}_r^{(d-1)}\right]_{d}
        \end{pmatrix} = \begin{pmatrix}
            \left[\widetilde{\mathbf{T}}_r^{(2)}\right]_{1} + \left[\widetilde{\mathbf{T}}_r^{(1)}\right]_{2}\\[6pt]
            \left[\widetilde{\mathbf{T}}_r^{(3)}\right]_{1} + \left[\widetilde{\mathbf{T}}_r^{(1)}\right]_{3}\\[6pt]
            \vdots\\[6pt]
            \left[\widetilde{\mathbf{T}}_r^{(d)}\right]_{1} + \left[\widetilde{\mathbf{T}}_r^{(1)}\right]_{d}\\[6pt]
            \left[\widetilde{\mathbf{T}}_r^{(3)}\right]_{2} + \left[\widetilde{\mathbf{T}}_r^{(2)}\right]_{3}\\[6pt]
            \vdots\\[6pt]
            \left[\widetilde{\mathbf{T}}_r^{(d)}\right]_{d-1} + \left[\widetilde{\mathbf{T}}_r^{(d-1)}\right]_{d}
        \end{pmatrix}
    \end{equation}
    An index mapping function is available for indexing row vectors of $\mathbf{T}_r$ in the form of
    \begin{equation}
        [\mathbf{T}_r]_{\tau(i,j), \cdot} =  \left[\widetilde{\mathbf{T}}_r^{(i)}\right]_{j} + \left[\widetilde{\mathbf{T}}_r^{(j)}\right]_{i}
    \end{equation}
    and for the $(\tau(i,j), k)$-th element of $\mathbf{T}_r$, we write
    \begin{equation}
        [\mathbf{T}_r]_{\tau(i,j), k} = \left\{\left[\widetilde{\mathbf{T}}_r^{(i)}\right]_{j}\right\}_k + \left\{\left[\widetilde{\mathbf{T}}_r^{(j)}\right]_{i}\right\}_k = \left[\widetilde{\mathbf{T}}_r^{(i)}\right]_{j,k} + \left[\widetilde{\mathbf{T}}_r^{(j)}\right]_{i,k}
    \end{equation}
    Therefore, we will have
    \begin{equation}
    \begin{aligned}
         \left[\mathbf{T}_r v_{\mathbf{X}, t}\right]_{\tau(i,j)} &= \sum_{\tau(m,n)=1}^{\frac{d(d-1)}{2}} [\mathbf{T}_r]_{\tau(i,j), \tau(m,n)} [v_{\mathbf{X},t}]_{\tau(m,n)}\\
         &= \sum_{\tau(m,n)=1}^{\frac{d(d-1)}{2}} \left(\left[\widetilde{\mathbf{T}}_r^{(i)}\right]_{j,\tau(m,n)} + \left[\widetilde{\mathbf{T}}_r^{(j)}\right]_{i,\tau(m,n)}\right)[v_{\mathbf{X},t}]_{\tau(m,n)}\\
         &= \sum_{\substack{u\in N_r(i)\\u\neq j}} w_{iu} X_{u,t}X_{j,t} + \sum_{\substack{v\in N_r(j)\\v\neq i}} w_{jv} X_{i,t}X_{v,t}
    \end{aligned}
    \end{equation}
    where the last step is because of \eqref{proof eqn: thm 2.8 eqn 3} and \eqref{proof eqn: thm 2.8 eqn 4}. Therefore, we have successfully shown the existence of such linear transformation $\mathbf{T}_r$ depending on the connection weight matrix $\mathbf{W}$ and stage of neighours $r$.

    To aid understanding, we implement the above proof step-by-step in Example \ref{appendix example}.
\end{proof}

\begin{example}
\label{appendix example}
    Let's consider a simple 4 node network example, illustrated in Figure \ref{fig: appendix example}, and for $r=1$ we are able to write $N_1(1) = \{2,3,4\}, N_1(2) = \{1,4\}, N_1(3) = \{1,4\}$ and $N_1(4) = \{1,2,3\}$.
    \begin{figure}[ht]
        \centering
        \includegraphics[width=.6\linewidth]{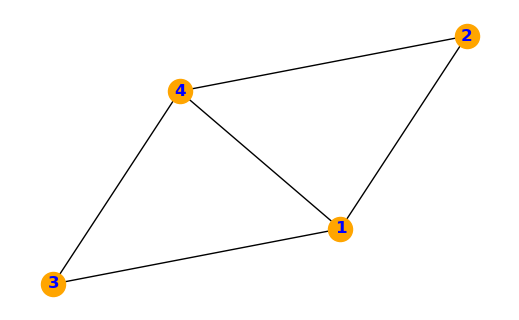}
        \caption{4 node network for this toy example}
        \label{fig: appendix example}
    \end{figure}
    
    We first consider $\sum_{\substack{u\in N_1(i)\\u\neq j}} w_{iu} X_{u,t}X_{j,t}$ for different values of $i$.
    
    For $i=1$, we will have
    \begin{equation}
    \label{appendix eqn: example}
    \begin{cases}
        w_{13}X_{2,t}X_{3,t} + w_{14}X_{2,t}X_{4,t} = w_{13}[v_{\mathbf{X},t}]_{(3,2)} + w_{14}[v_{\mathbf{X},t}]_{(4,2)} &\hbox{if $j=2$}\\
        w_{12}X_{2,t}X_{3,t} + w_{14}X_{3,t}X_{4,t} = w_{12}[v_{\mathbf{X},t}]_{(3,2)} + w_{14}[v_{\mathbf{X},t}]_{(4,3)} &\hbox{if $j=3$}\\
        w_{12}X_{2,t}X_{4,t} + w_{13}X_{3,t}X_{4,t} = w_{12}[v_{\mathbf{X},t}]_{(4,2)} + w_{13}[v_{\mathbf{X},t}]_{(4,3)} &\hbox{if $j=4$}
    \end{cases}
    \end{equation}
    which matches \eqref{proof eqn: thm 2.8 eqn 1} along with $i=1$ and $j=2,3,4$ respectively. Hence, for $r=1$ and $i=1$, we can write $\widetilde{\mathbf{T}}_1^{(1)}$ as
    \begin{equation}
        \widetilde{\mathbf{T}}_1^{(1)} = \begin{pmatrix}
            0 & 0 & 0 & w_{13} & w_{14} & 0\\
            0 & 0 & 0 & w_{12} & 0 & w_{14}\\
            0 & 0 & 0 & 0 & w_{12} & w_{13}
        \end{pmatrix} = \begin{pmatrix}
            \left[\widetilde{\mathbf{T}}_1^{(1)}\right]_{2}\\[6pt]
            \left[\widetilde{\mathbf{T}}_1^{(1)}\right]_{3}\\[6pt]
            \left[\widetilde{\mathbf{T}}_1^{(1)}\right]_{4}
        \end{pmatrix}
    \end{equation}
    so that $\widetilde{\mathbf{T}}_1^{(1)}v_{\mathbf{X},t}$ matches \eqref{appendix eqn: example}. In particular, the explicit form using tuple index is
    \begin{equation}
        \widetilde{\mathbf{T}}_1^{(1)}v_{\mathbf{X},t} = \begin{pmatrix}
            0 & 0 & 0 & w_{13} & w_{14} & 0\\
            0 & 0 & 0 & w_{12} & 0 & w_{14}\\
            0 & 0 & 0 & 0 & w_{12} & w_{13}
        \end{pmatrix} \begin{pmatrix}
            [v_{\mathbf{X},t}]_{(2,1)}\\
            [v_{\mathbf{X},t}]_{(3,1)}\\
            [v_{\mathbf{X},t}]_{(4,1)}\\
            [v_{\mathbf{X},t}]_{(3,2)}\\
            [v_{\mathbf{X},t}]_{(4,2)}\\
            [v_{\mathbf{X},t}]_{(4,3)}\\
        \end{pmatrix} = \begin{pmatrix}
            w_{13}[v_{\mathbf{X},t}]_{(3,2)} + w_{14}[v_{\mathbf{X},t}]_{(4,2)}\\
            w_{12}[v_{\mathbf{X},t}]_{(3,2)} + w_{14}[v_{\mathbf{X},t}]_{(4,3)}\\
            w_{12}[v_{\mathbf{X},t}]_{(4,2)} + w_{13}[v_{\mathbf{X},t}]_{(4,3)}
        \end{pmatrix}
    \end{equation}
    Similarly we can also write for other nodes $i=2,3,4$ that
    \begin{align}
        \widetilde{\mathbf{T}}_1^{(2)} &= \begin{pmatrix}
            0 & 0 & w_{24} & 0 & 0 & 0\\
            0 & w_{21} & 0 & 0 & 0 & w_{24}\\
            0 & 0 & w_{21} & 0 & 0 & 0
        \end{pmatrix} = \begin{pmatrix}
            \left[\widetilde{\mathbf{T}}_1^{(2)}\right]_{1}\\[6pt]
            \left[\widetilde{\mathbf{T}}_1^{(2)}\right]_{3}\\[6pt]
            \left[\widetilde{\mathbf{T}}_1^{(2)}\right]_{4}
        \end{pmatrix}\\
        \widetilde{\mathbf{T}}_1^{(3)} &= \begin{pmatrix}
            0 & 0 & w_{34} & 0 & 0 & 0\\
            w_{31} & 0 & 0 & 0 & w_{34} & 0\\
            0 & 0 & w_{31} & 0 & 0 & 0
        \end{pmatrix} = \begin{pmatrix}
            \left[\widetilde{\mathbf{T}}_1^{(3)}\right]_{1}\\[6pt]
            \left[\widetilde{\mathbf{T}}_1^{(3)}\right]_{2}\\[6pt]
            \left[\widetilde{\mathbf{T}}_1^{(3)}\right]_{4}
        \end{pmatrix}\\
        \widetilde{\mathbf{T}}_1^{(4)} &= \begin{pmatrix}
            w_{42} & w_{43} & 0 & 0 & 0 & 0\\
            w_{41} & 0 & 0 & w_{43} & 0 & 0\\
            0 & w_{41} & 0 & w_{42} & 0 & 0
        \end{pmatrix} = \begin{pmatrix}
            \left[\widetilde{\mathbf{T}}_1^{(4)}\right]_{1}\\[6pt]
            \left[\widetilde{\mathbf{T}}_1^{(4)}\right]_{2}\\[6pt]
            \left[\widetilde{\mathbf{T}}_1^{(4)}\right]_{3}
        \end{pmatrix}
    \end{align}
    Finally, our desired $\mathbf{T}_1$ matrix (again, in this example we fix $r=1$) will be the sum of two parts, and each part is constructed from $\widetilde{\mathbf{T}}_1^{(1)}, \widetilde{\mathbf{T}}_1^{(2)}, \widetilde{\mathbf{T}}_1^{(3)}$ and $\widetilde{\mathbf{T}}_1^{(4)}$. Specifically,
    \begin{equation}
    \begin{aligned}
        \mathbf{T}_1 &= \begin{pmatrix}
            \left[\widetilde{\mathbf{T}}_1^{(2)}\right]_{1}\\[6pt]
            \left[\widetilde{\mathbf{T}}_1^{(3)}\right]_{1}\\[6pt]
            \left[\widetilde{\mathbf{T}}_1^{(4)}\right]_{1}\\[6pt]
            \left[\widetilde{\mathbf{T}}_1^{(3)}\right]_{2}\\[6pt]
            \left[\widetilde{\mathbf{T}}_1^{(4)}\right]_{2}\\[6pt]
            \left[\widetilde{\mathbf{T}}_1^{(4)}\right]_{3}
        \end{pmatrix} + \begin{pmatrix}
            \left[\widetilde{\mathbf{T}}_1^{(1)}\right]_{2}\\[6pt]
            \left[\widetilde{\mathbf{T}}_1^{(1)}\right]_{3}\\[6pt]
            \left[\widetilde{\mathbf{T}}_1^{(1)}\right]_{4}\\[6pt]
            \left[\widetilde{\mathbf{T}}_1^{(2)}\right]_{3}\\[6pt]
            \left[\widetilde{\mathbf{T}}_1^{(2)}\right]_{4}\\[6pt]
            \left[\widetilde{\mathbf{T}}_1^{(3)}\right]_{4}
        \end{pmatrix}\\
        &= \begin{pmatrix}
            0 & 0 & w_{24} & 0 & 0 & 0\\
            0 & 0 & w_{34} & 0 & 0 & 0\\
            w_{42} & w_{43} & 0 & 0 & 0 & 0\\
            w_{31} & 0 & 0 & 0 & w_{34} & 0\\
            w_{41} & 0 & 0 & w_{43} & 0 & 0\\
            0 & w_{41} & 0 & w_{42} & 0 & 0
        \end{pmatrix} + \begin{pmatrix}
            0 & 0 & 0 & w_{13} & w_{14} & 0\\
            0 & 0 & 0 & w_{12} & 0 & w_{14}\\
            0 & 0 & 0 & 0 & w_{12} & w_{13}\\
            0 & w_{21} & 0 & 0 & 0 & w_{24}\\
            0 & 0 & w_{21} & 0 & 0 & 0\\
            0 & 0 & w_{31} & 0 & 0 & 0
        \end{pmatrix}\\
        &= \begin{pmatrix}
            0 & 0 & w_{24} & w_{13} & w_{14} & 0\\
            0 & 0 & w_{34} & w_{12} & 0 & w_{14}\\
            w_{42} & w_{43} & 0 & 0 & w_{12} & w_{13}\\
            w_{31} & w_{21} & 0 & 0 & w_{34} & w_{24}\\
            w_{41} & 0 & w_{21} & w_{43} & 0 & 0\\
            0 & w_{41} & w_{31} & w_{42} & 0 & 0
        \end{pmatrix}
    \end{aligned}
    \end{equation}
\end{example}

\begin{proof}
    This is the proof of the conditional unbiasedness of squared return as the conditional variance proxy in our GNGARCH, stated in Section \ref{subsubsec: volatility/variance proxy}, with $\widehat{\Sigma}_t$ as the model conditional variance matrix.

    By \eqref{eqn: GNGARCH equation 1} and the SWN process $\{\mathbf{Z}_{t}\}$ with zero mean and unit variance, together with the $\mathcal{F}_{t-1}$-measurability of (elementary) model variance $\hat{\sigma}_{i,t}^2$ and model covariance $\hat{\sigma}_{ij,t}$ (which are shown in the previous proof of $\{\eta_{i,t}\}$ and $\{\eta_{ij,t}\}$ being as white noise (WN) processes), we can therefore say that the model conditional variance matrix $\widehat{\Sigma}_t$ is also $\mathcal{F}_{t-1}$-measurable, then
    \begin{align}
        \mathbb{E}(\mathbf{X}_t\mathbf{X}_t^T\mid\mathcal{F}_{t-1}) &= \mathbb{E}\left(\widehat{\Sigma}_t^{1/2}\mathbf{Z}_t\mathbf{Z}_t^T \left(\widehat{\Sigma}_t^{1/2}\right)^T \mid \mathcal{F}_{t-1}\right)\\
        &= \widehat{\Sigma}_t^{1/2}\mathbb{E}\left(\mathbf{Z}_t\mathbf{Z}_t^T \mid \mathcal{F}_{t-1}\right)\left(\widehat{\Sigma}_t^{1/2}\right)^T\\
        &= \widehat{\Sigma}_t^{1/2}\mathbb{E}\left(\mathbf{Z}_t\mathbf{Z}_t^T\right)\left(\widehat{\Sigma}_t^{1/2}\right)^T = \widehat{\Sigma}_t
    \end{align}
\end{proof}

\begin{proof}
    This is the proof of the statement in Section \ref{subsubsec: volatility/variance proxy} that using the model conditional variance but not the model standard deviation. For simplicity, we only discuss the univariate analogy here, and a multivariate study is similar to the univariate case.

    Consider the univariate case where the squared return $X_t^2$ is a conditional unbiased estimator of model conditional variance $\sigma_t^2$ that
    \begin{equation}
        \mathbb{E}(X_t^2 \mid \mathcal{F}_{t-1}) = \sigma_t^2 \iff \sqrt{\mathbb{E}(X_t^2 \mid \mathcal{F}_{t-1})} = \sigma_t
    \end{equation}
    However, by Jensen's inequality, since $f(x) = \sqrt{x}$ is a concave function, we shall have
    \begin{equation}
        \mathbb{E}\left(\sqrt{X_t^2} \mid \mathcal{F}_{t-1}\right) = \mathbb{E}(|X_t| \mid \mathcal{F}_{t-1}) \leq \sqrt{\mathbb{E}(X_t^2 \mid \mathcal{F}_{t-1})} = \sigma_t
    \end{equation}
    This means that the absolute return $|X_t|$ is a conditionally biased estimate of the conditional standard deviation $\sigma_t$.
\end{proof}

\begin{proof}
This is the proof of the invariance between the (averaged) QLIKE \eqref{eqn: QLIKE} and NLL \eqref{eqn: NLL} up to additive constants and an overall scale factor, stated in Section \ref{subsubsec: loss function}. 

Recall that the joint density of observed returns (excluding known initial return $\mathbf{X}_0$) can be written in terms of
\begin{equation}
    p(\mathbf{X}_1,\cdots, \mathbf{X}_{T-1}) = p(\mathbf{X}_{T-1}\mid\mathcal{F}_{T-2})\cdots p(\mathbf{X}_1\mid\mathcal{F}_0) = \prod_{t=1}^{T-1} p(\mathbf{X}_t\mid\mathcal{F}_{t-1}) = \prod_{t=1}^{T-1} \mathcal{N}(\mathbf{X}_t; 0, \widehat{\Sigma}_t)
\end{equation}
Then the (average) NLL for the joint density is, for multivariate normal distributions,
\begin{equation}
\begin{aligned}
    L_{\text{NLL}}(\mathbf{X}_t\mathbf{X}_t^T, \widehat{\Sigma}_t) &= -\frac{1}{T-1}\sum_{t=1}^{T-1} \log \mathcal{N}(\mathbf{X}_t; 0, \widehat{\Sigma}_t)\\
    &= \frac{1}{T-1}\sum_{t=1}^{T-1} \left(\frac{d}{2}\log(2\pi) + \frac{1}{2}\log |\widehat{\Sigma}_t| + \frac{1}{2} \mathbf{X}_t^T \widehat{\Sigma}_t^{-1}\mathbf{X}_t\right)\\
    &= \frac{1}{2}L_{\text{QLIKE}}(\mathbf{X}_t\mathbf{X}_t^T, \widehat{\Sigma}_t) + \frac{d}{2}\log(2\pi)
\end{aligned}
\end{equation}
\end{proof}

\section{Additional Figures and Tables}

\begin{figure}[ht]
    \centering
    \includegraphics[width=\linewidth]{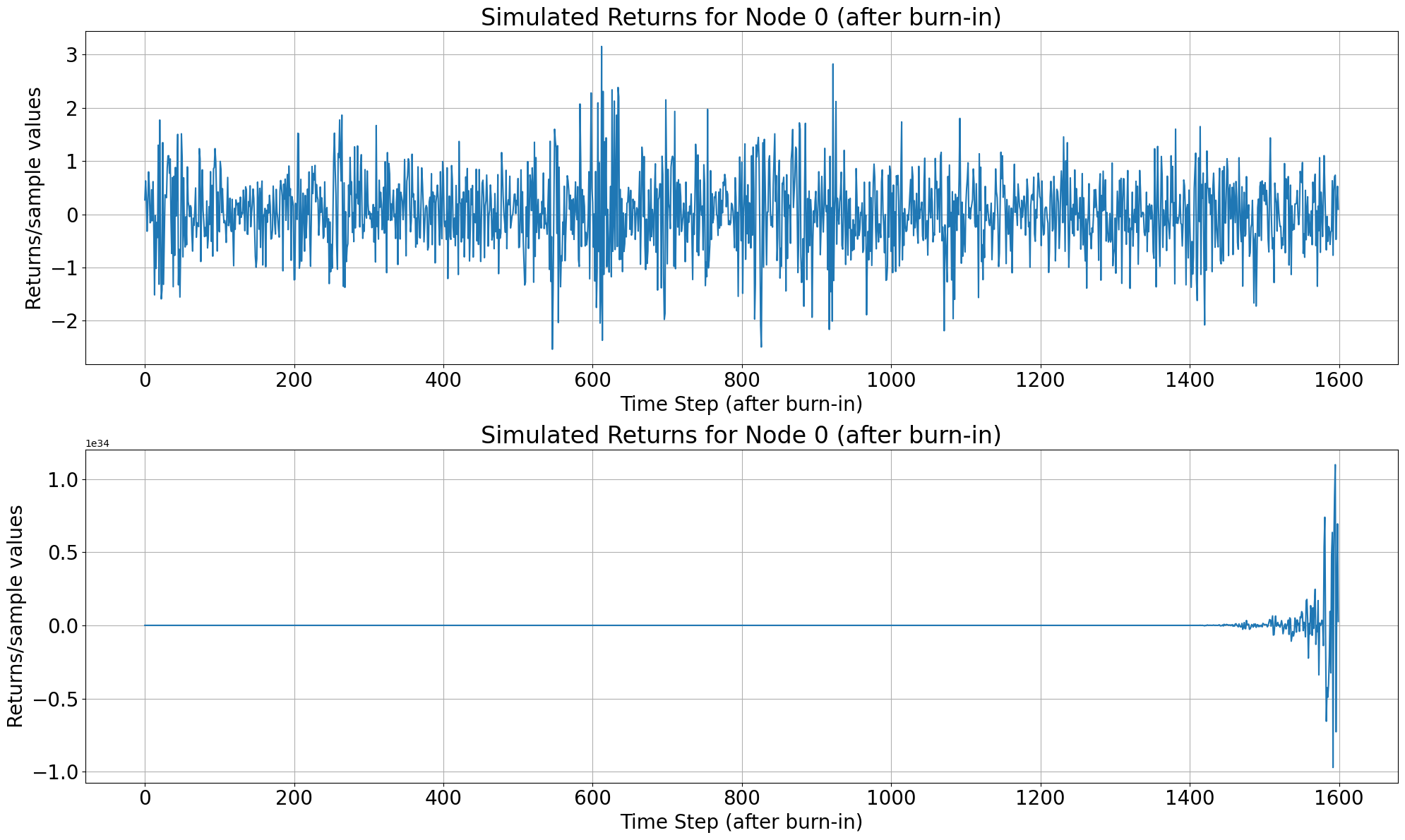}
    \caption{Simulation of $\{X_{0,t}\}$ on the example network showing in Figure \ref{fig: simulation network}, generated under the simulation scheme described in Section \ref{subsec: model simulation and validity}. A total of 2000 samples are drawn under random seed 0, and we discard the first 400 samples as burnin. Top panel: a convergent and stationary simulation of $\{X_{0,t}\}$ using model parameters $(\alpha_0, \alpha_1, \gamma_1, \beta_{11}, \delta_{11}) = (0.05, 0.20, 0.60, 0.05, 0.05)$, with $\alpha_1+\gamma_1+\beta_{11}+\delta_{11}=0.90<1$ satisfying Conjecture \ref{conj: sufficient stationarity}. Bottom panel: a divergent and explosive simulation of $\{X_{0,t}\}$ using model parameters $(\alpha_0, \alpha_1, \gamma_1, \beta_{11}, \delta_{11}) = (0.06, 0.40, 0.55, 0.10, 0.10)$,  with $\alpha_1+\gamma_1+\beta_{11}+\delta_{11}=1.05>1$ rejecting Conjecture \ref{conj: sufficient stationarity}.}
    \label{fig: convergent-divergent simulation}
\end{figure}

\begin{table}[ht]
\centering
\begin{tabular}{c c}
    \hline
    Model (of $d$ nodes/dimension $d$) & Active parameters\\
    \hline
    GNGARCH($1,1,[1],[1]$) & 5\\
    VARMA($1,1$), no intercept & $2d^2$\\
    VARMA($1,1$), with intercept & $2d^2 + d$\\
    \hline
\end{tabular}
\caption{Comparison of model parameters between GNGARCH and VARMA.}
\label{tab: model parameter number comparison}
\end{table}

\begin{table}[ht]
\centering
\begin{tabular}{l c r r}
    \hline
    Parameter & True  & Estimate (MSE) & Estimate (NLL)\\
    \hline
    $\alpha_0$ & 0.05 & 0.08 $\pm$ 0.03 & 0.04 $\pm$ 0.02\\
    $\alpha_1$ & 0.20 & 0.15 $\pm$ 0.05 & 0.21 $\pm$ 0.07\\
    $\gamma_1$ & 0.60 & 0.51 $\pm$ 0.13 & 0.56 $\pm$ 0.13\\
    $\beta_{11}$ & 0.05 & 0.05 $\pm$ 0.04 & 0.03 $\pm$ 0.02\\
    $\delta_{11}$ & 0.05 & 0.07 $\pm$ 0.03 & 0.08 $\pm$ 0.07\\
    \hline
\end{tabular}
\caption{Parameter estimates up to 2 decimal places, in the form of: sample mean $\pm$ sample standard deviation. Note that since our simulation network is rather simple, and simulated returns from the true parameters have low probabilities of outliers than the real return series, both MSE and NLL show good fitting results over different seeds from 0 to 19. For some parameters ($\alpha_1, \gamma_1, \delta_{11}$) MSE provides estimation with smaller sample variance (for $\gamma_1$, the raw sample variance without rounding is smaller for MSE), while NLL outperforms MSE on the remaining parameters.}
\label{tab: iterative parameter fitting statistics}
\end{table}

\begin{figure}[ht]
    \centering
    \includegraphics[width=\linewidth]{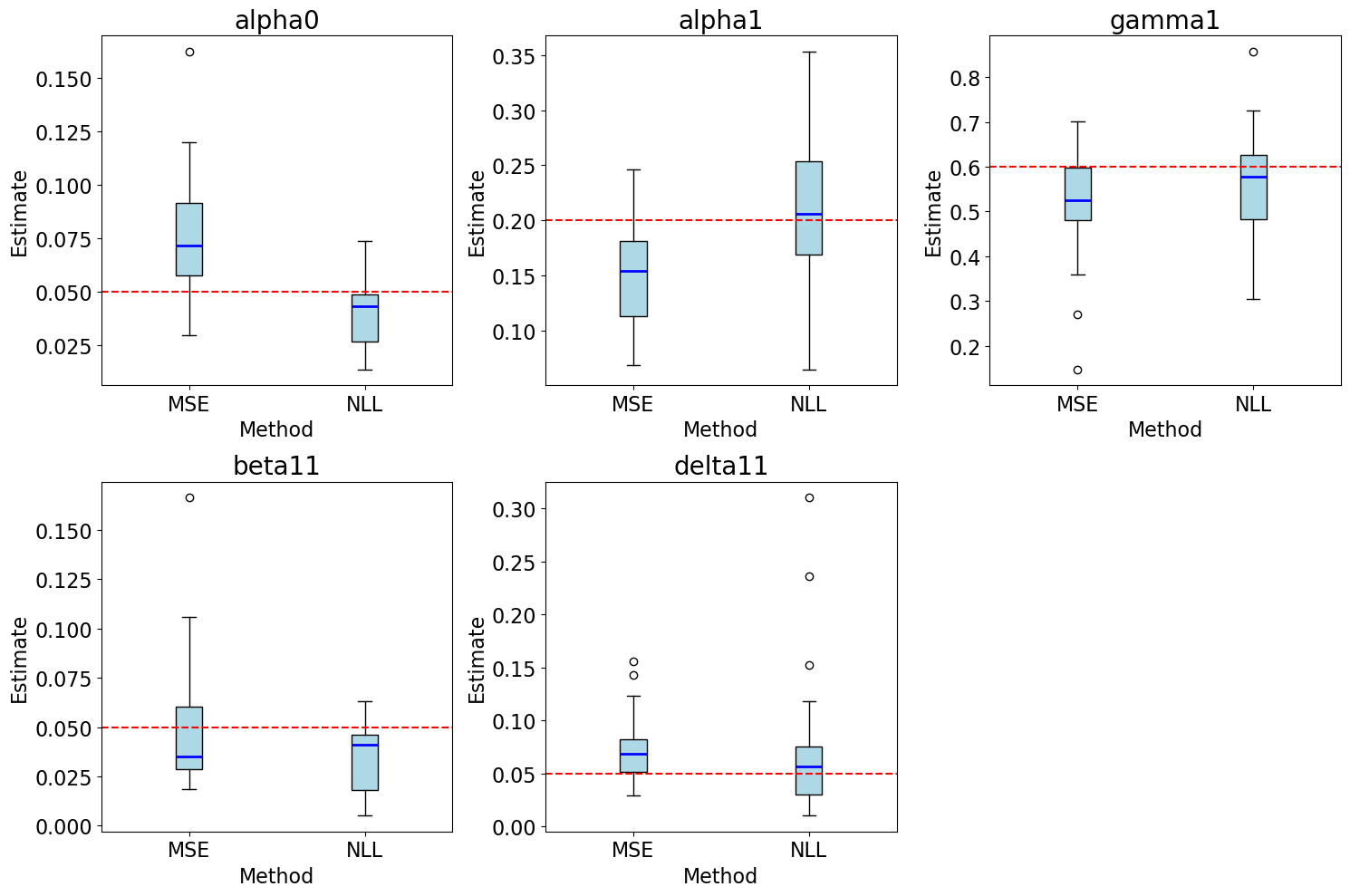}
    \caption{Boxplots for each parameter estimates using different simulated data by different random seeds from 0 to 19. Despite of mean and standard deviation in Table \ref{tab: iterative parameter fitting statistics}, the above boxplots show optimisation via NLL can give estimated parameter median much closer to the true value than MSE.}
    \label{fig: boxplot parameter fitting}
\end{figure}

\begin{table}[ht]
\centering
\begin{tabular}{l r r r}
    \hline
    Parameter & MSE (75 stocks) & MSE (47 stocks) & \% change\\
    \hline
    $\alpha_0$ & 0.0232 & 0.0228 & $-1.72$\\
    $\alpha_1$ & 0.0702 & 0.0699 & $-0.43$\\
    $\gamma_1$ & 0.1693 & 0.1660 & $-1.95$\\
    $\beta_{11}$ & 0.0236 & 0.0229 & $-2.97$\\
    $\delta_{11}$ & 0.0432 & 0.0423 & $-2.08$\\
    \hline
\end{tabular}
\caption{Parameter estimates by using MSE on the dataset containing all 75 stocks and filtered 47 stocks, results up to 4 decimal places, with the percentage change up to 2 decimal places.
Note how close the results of the MSE parameter estimation between total 75 stocks and filtered 47 stocks with small parameter percentage change.}
\label{tab: stock fitting parameter results comparison, MSE}
\end{table}

\begin{table}[ht]
\centering
\begin{tabular}{l r r r}
    \hline
    Parameter & NLL (75 stocks) & NLL (47 stocks) & \% change \\
    \hline
    $\alpha_0$ & 0.0005 & 0.0003 & $-40.00$\\
    $\alpha_1$ & 0.1648 & 0.1401 & $-14.99$\\
    $\gamma_1$ & 0.7072 & 0.7626 & $7.83$\\
    $\beta_{11}$ & 0.0008 & 0.0003 & $-62.50$\\
    $\delta_{11}$ & 0.0039 & 0.0026 & $-33.33$\\
    \hline
\end{tabular}
\caption{Parameter estimates by using NLL on the dataset containing all 75 stocks and filtered 47 stocks, results up to 4 decimal places, with the percentage change up to 2 decimal places. Compared with percentage change for MSE estimate in Table \ref{tab: stock fitting parameter results comparison, MSE}, here we observe significantly larger percentage change for NLL-based parameter estimates.}
\label{tab: stock fitting parameter results comparison, NLL}
\end{table}

\begin{figure}[ht]
    \centering
    \includegraphics[width=\linewidth]{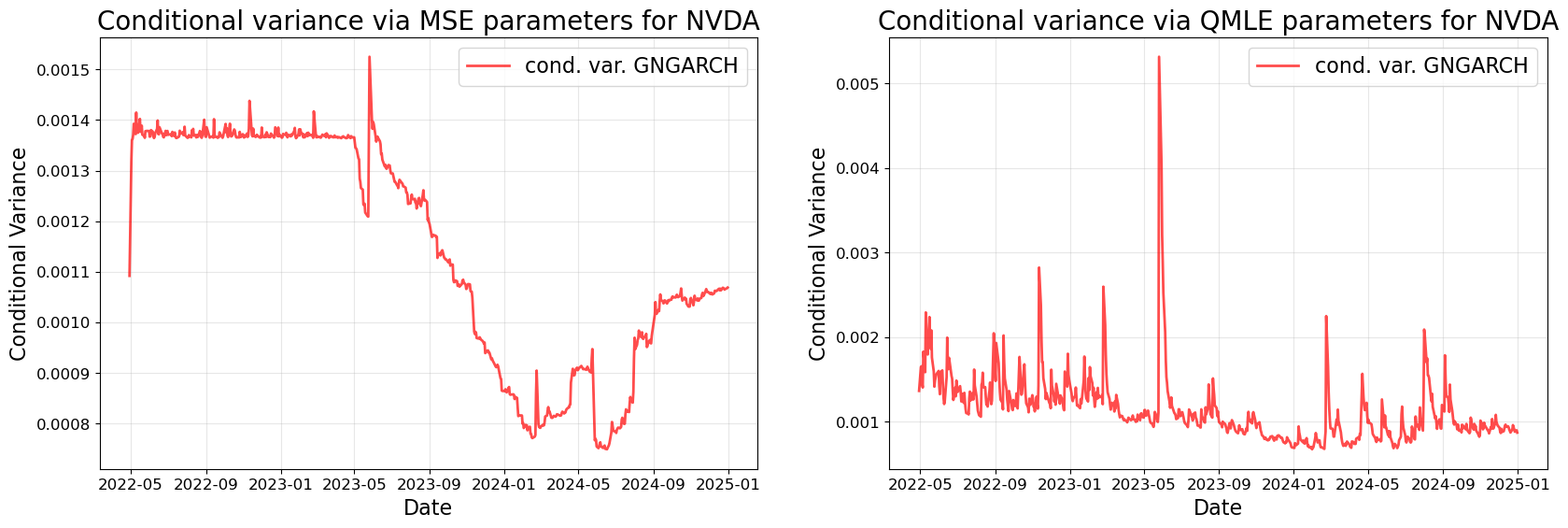}
    \caption{The fitted GNGARCH($1,1,[1],[1]$) model with rescaled conditional variance by using MSE-based parameters (left) and QLIKE/NLL-based parameters, which are indeed the QMLE (right).}
    \label{fig: NVDA training volatility}
\end{figure}

\begin{figure}[ht]
    \centering
    \includegraphics[width=\linewidth]{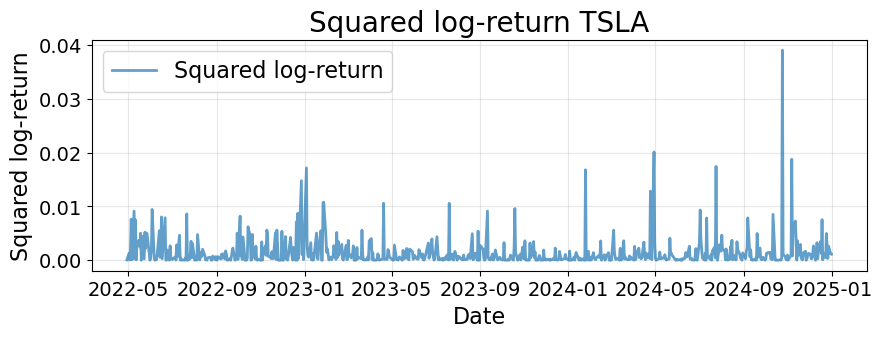}
    \caption{Squared log-returns over time for stock TSLA.}
    \label{fig: squared log-returns TSLA}
\end{figure}

\begin{figure}[ht]
    \centering
    \includegraphics[width=\linewidth]{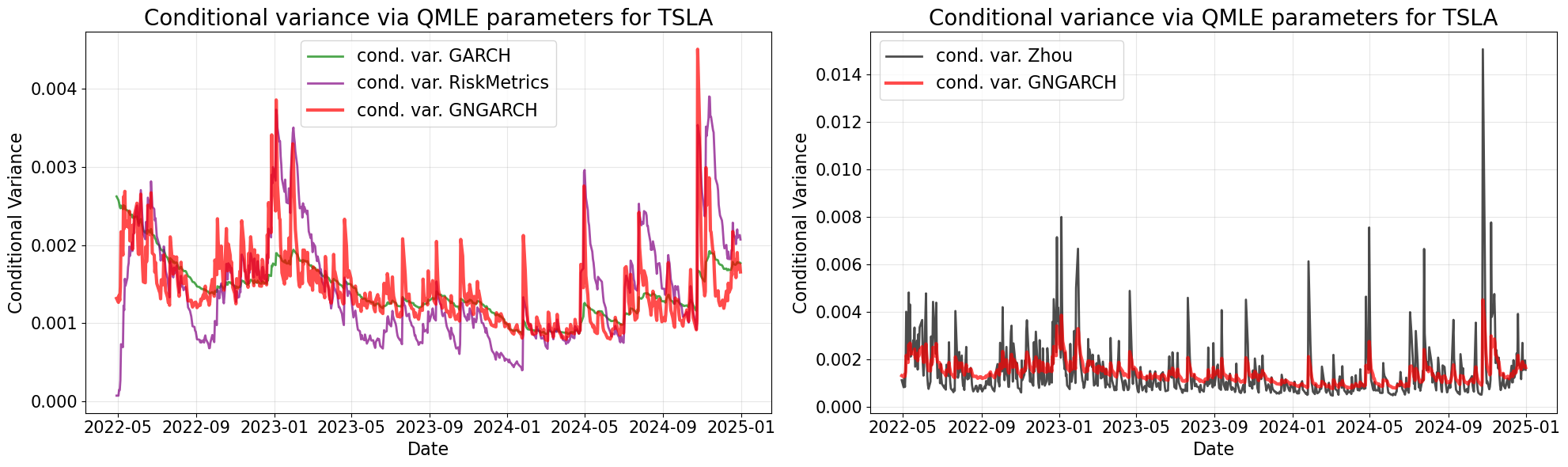}
    \caption{Conditional variance comparison for TSLA across models. Left: comparison between the best-fit univariate GARCH (green), RiskMetrics ($\lambda=0.94$, purple) and the fitted GNGARCH$(1,1,[1],[1])$ (thick red). Right: comparison between Zhou’s network GARCH with QMLE-fitted parameters (black) and the fitted GNGARCH$(1,1,[1],[1])$ (thick red).}
    \label{fig: TSLA volatility comparison}
\end{figure}

\begin{figure}[ht]
    \centering
    \includegraphics[width=\linewidth]{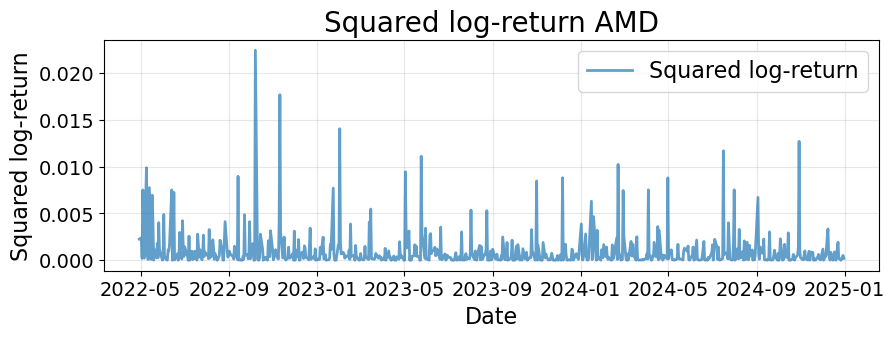}
    \caption{Squared log-returns over time for stock AMD.}
    \label{fig: squared log-returns AMD}
\end{figure}

\begin{figure}[ht]
    \centering
    \includegraphics[width=\linewidth]{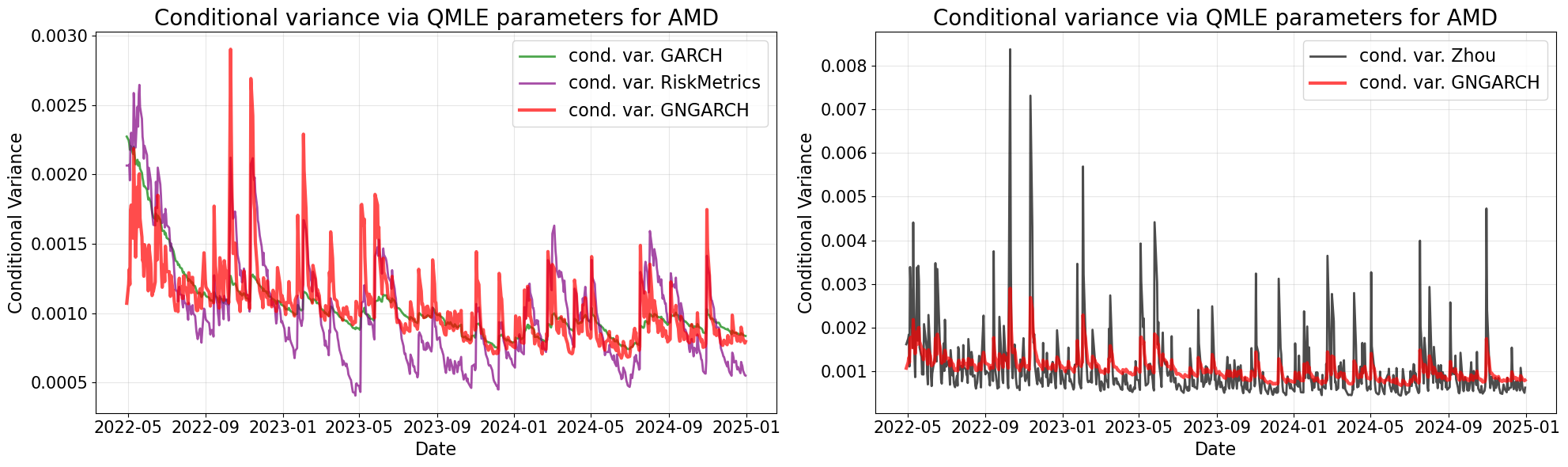}
    \caption{Conditional variance comparison for AMD across models. Left: comparison between the best-fit univariate GARCH (green), RiskMetrics ($\lambda=0.94$, purple) and the fitted GNGARCH$(1,1,[1],[1])$ (thick red). Right: comparison between Zhou’s network GARCH with QMLE-fitted parameters (black) and the fitted GNGARCH$(1,1,[1],[1])$ (thick red).}
    \label{fig: AMD volatility comparison}
\end{figure}

\begin{figure}[ht]
    \centering
    \includegraphics[width=\linewidth]{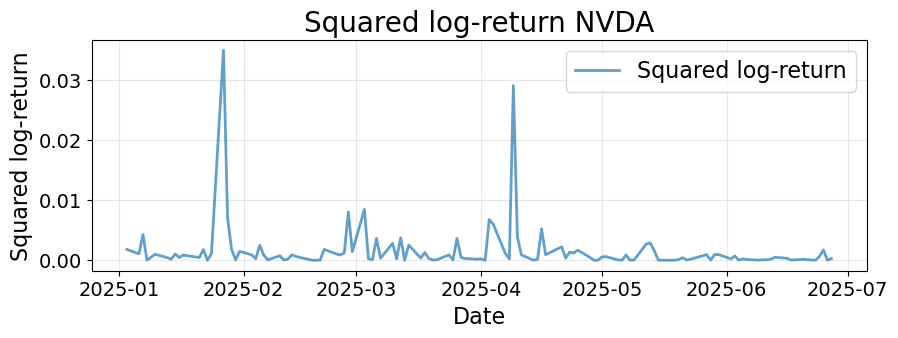}
    \caption{Squared log-returns for stock NVDA on the validation set, from 3rd January, 2025 to 27th June, 2025.}
    \label{fig: squared log-returns NVDA val}
\end{figure}

\begin{figure}[ht]
    \centering
    \includegraphics[width=\linewidth]{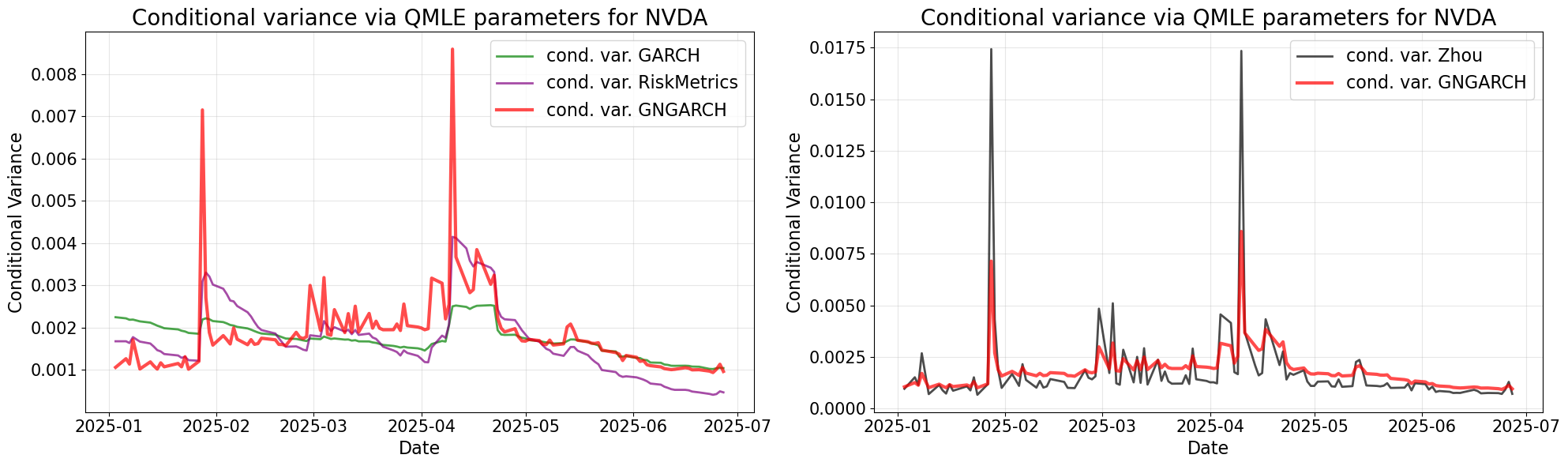}
    \caption{Conditional variance comparison for NVDA by different volatility models, on the validation set from 3rd January, 2025 to 27th June, 2025.}
    \label{fig: NVDA volatility comparison val}
\end{figure}

\begin{figure}[ht]
    \centering
    \includegraphics[width=\linewidth]{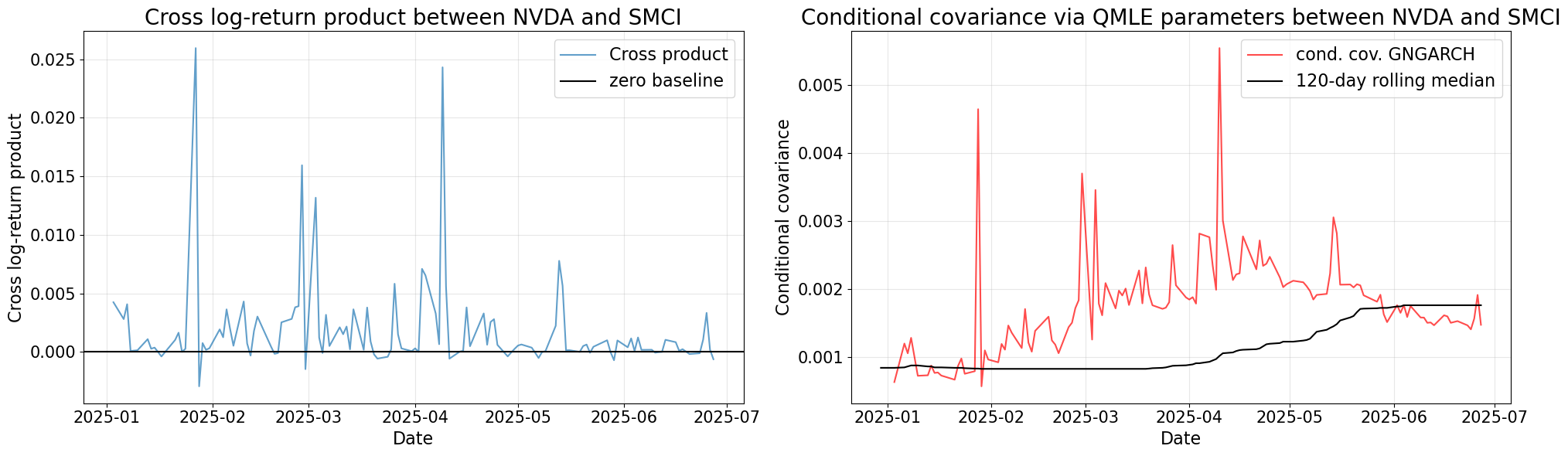}
    \caption{Left: Cross-product of log-returns over time between stock NVDA and SMCI. Right: conditional covariance between NVDA and SMCI, plotted against its (120-day) rolling median. Conditional covariances above the base rolling median indicate the same direction co-movement, vice versa.}
    \label{fig: NVDA-SMCI covolatility val}
\end{figure}


\end{document}